\documentclass[12pt,preprint]{aastex}
\usepackage{natbib,graphicx}

\shorttitle{Extended radio emission in MOJAVE Blazars}
\shortauthors{Kharb et al.}

\begin{document}

\title{Extended radio emission in MOJAVE Blazars: Challenges to
Unification}

\author{P. Kharb\footnote{$^*$Major part of this work completed at
Department of Physics, Purdue University, 525 Northwestern Ave,
West Lafayette, IN 47907}\altaffilmark{1,2*}}
\affil{Dept of Physics, Rochester Institute of Technology, Rochester, NY 14623}
\email{kharb@cis.rit.edu}
\author{M. L. Lister\altaffilmark{2}}
\affil{Dept of Physics, Purdue University, 525 Northwestern Ave,
West Lafayette, IN 47907}
\and \author{N. J. Cooper\altaffilmark{2}}

\begin{abstract}
We present the results of a study on the kiloparsec-scale radio emission 
in the complete flux density limited MOJAVE sample, comprising 135 radio-loud 
AGNs. New 1.4~GHz VLA radio images of six quasars, and previously unpublished 
images of 21 blazars, are presented, along with an analysis of the high 
resolution (VLA A-array) 1.4~GHz emission for the entire sample. While 
extended emission is detected in the majority of the sources, about 7\% 
of the sources exhibit only radio core emission. 
We expect more sensitive radio observations, however, to detect faint emission in 
these sources, as we have detected in the erstwhile ``core-only'' source, 1548+056.
The kiloparsec-scale radio morphology varies widely across the sample. 
Many BL Lacs exhibit extended radio power and kiloparsec-scale morphology 
typical of powerful FRII jets, while a substantial number of
quasars possess radio powers intermediate between FRIs and FRIIs. This poses 
challenges to the simple radio-loud unified scheme, which links BL~Lacs to FRIs
and quasars to FRIIs. We find a significant 
correlation between extended radio emission and parsec-scale jet speeds: the 
more radio powerful sources possess faster jets. 
{This indicates that the 1.4 GHz (or low frequency) radio emission is 
indeed related to jet kinetic power.} Various properties 
such as extended radio power and apparent parsec-scale jet speeds vary 
smoothly between different blazar subclasses, suggesting that,
{at least in terms of radio jet properties}, the distinction 
between quasars and BL~Lac objects, at an emission-line equivalent width of 
$5\AA$ is essentially an arbitrary one.
{While the two blazar subclasses display a smooth continuation in properties,
they often reveal differences in the correlation test results, when
considered separately. This can be understood if, unlike quasars, BL~Lacs do 
not constitute a homogeneous population, but rather include both FRI and FRII 
radio galaxies for their parent population.}
{It could also just be due to small number statistics.}
We find that the ratio of the radio core luminosity to the $k$-corrected 
optical luminosity ($R_v$) appears to be a better indicator of orientation 
for this blazar sample, than the traditionally used radio core prominence 
parameter ($R_c$). {Based on the assumption that the extended radio 
luminosity is} affected by 
the kiloparsec-scale environment, we define the ratio of extended radio power 
to absolute optical magnitude ($L_{ext}/M_{abs}$) as a proxy for 
environmental effects. Trends with this parameter suggest that the parsec-scale 
jet speeds {and the parsec-to-kiloparsec jet misalignments} are not affected 
by the large-scale environment, but are more likely to depend upon factors 
intrinsic to the AGN, or its local parsec-scale environment. 
{The jet speeds could, for instance, be related to the black hole spins, while 
jet misalignments
could arise due to the presence of binary black holes, or kicks imparted to black
holes via black hole mergers, consistent both with radio morphologies 
resembling precessing jet models observed in some MOJAVE blazars, and the 
signature of a 90$\degr$ bump in the jet misalignment distribution,
attributed to low-pitch helical parsec-scale jets in the literature.
We suggest that some of the extremely misaligned MOJAVE blazar
jets could be ``hybrid'' morphology sources, with an FRI jet on one side and an
FRII jet on the other. Finally, it is tempting to speculate that environmental radio
boosting (as proposed for Cygnus A) could be responsible for blurring the
Fanaroff-Riley dividing line in the MOJAVE blazars, provided a substantial fraction
of them reside in dense (cluster) environments.}
\end{abstract}


\section{INTRODUCTION}
Radio-loud active galactic nuclei (AGNs) that are characterised by extreme 
variability in their radio cores, high and variable polarization, 
superluminal jet speeds and compact radio emission, are collectively 
referred to as `blazars' \citep{AngelStockman80}. Blazars comprise flat-spectrum 
radio-loud quasars and BL~Lac objects. Their extreme characteristics have commonly 
been understood to be a consequence of relativistic beaming effects due 
to a fast jet aligned close to our line of sight \citep{BlandfordKonigl79}.

The radio-loud unified scheme postulates that quasars and BL~Lacs are the 
beamed end-on counterparts of the Fanaroff-Riley \citep[FR,][]{FanaroffRiley74} 
type II and type I radio galaxies, respectively \citep{UrryPadovani95}. 
The relatively lower radio luminosity FRI radio galaxies are referred to as 
``edge-darkened'' since their brightest lobe emission lies closer to the 
radio cores and fades further out, while the higher radio luminosity 
FRII radio galaxies are ``edge-brightened'' because of the presence of bright 
and compact hot spots where the kpc-scale jets terminate. This definition 
turns out to be useful for FRI-FRII classification when hot spots are not 
clearly delineated ($cf.$ \S3.2).

The Fanaroff-Riley dichotomy has been proposed to arise due to 
differences in one or more of the following properties: host galaxy environment 
\citep{PrestagePeacock88} and jet-medium interaction \citep{Bicknell95}, 
jet composition \citep{Reynolds96}, black hole mass \citep{Ghisellini01}, black 
hole spin \citep{Meier99}, accretion rate and/or mode \citep{Baum95,Ghisellini01}. 
Many recent findings are, however, posing challenges for the standard 
unified scheme. These include the discovery of FRI quasars \citep{Heywood07},
FRII BL~Lacs \citep{Landt06}, and ``hybrid'' radio morphology sources with an 
FRI jet on one side of the core and an FRII jet on the other \citep{Gopal-Krishna00}.
One of the primary cited differences between quasars and BL~Lacs 
has been the presence of strong, broad emission lines in the quasar spectra
and their apparent absence in BL~Lacs. \citet{Stickel91} and \citet{Stocke91}
have defined the distinction between BL~Lacs and quasars at an emission line 
equivalent width of 5$\AA$. However, this distinction has been questioned 
\citep[e.g.,][]{ScarpaFalomo97,Urry99,Landt04}, and it is known that some 
BL~Lacs have a broad-line region 
\citep[e.g.,][]{Miller78,Stickel91,Vermeulen95,Corbett96}. 

In this paper, we examine the unified scheme and FR dichotomy in the 
MOJAVE\footnote{Monitoring Of Jets in Active galactic nuclei with VLBA 
Experiments. http://www.physics.purdue.edu/MOJAVE/} sample of blazars. 
MOJAVE is a long-term program to monitor radio brightness and 
polarization variations in the jets associated with active galaxies visible 
in the northern sky on parsec-scales with Very Long Baseline Interferometry 
(VLBI) \citep{Lister09}.
The {MOJAVE} sample consists of 135 sources satisfying the 
following criteria: (1) J2000.0 declination $>-20\degr$;
(2) galactic latitude $|b|>2.5\degr$;
(3) VLBA 2~cm correlated flux density exceeding 1.5 Jy (2 Jy for declination
south of $0\degr$) at any epoch between 1994.0 and 2004.0.
Of the 135 AGNs, 101 are classified as quasars, 22 as BL~Lac objects, 
and eight as radio galaxies of mostly the FRII-type. Four radio sources are 
yet to be identified on the basis of their emission line spectra and have no
redshift information. Overall, eight sources (four BL Lacs and four 
unidentified) have unknown or unreliable redshifts. 
Based on the synchrotron peak in the spectral energy distributions (SEDs),
BL Lacs have been divided into low, high, and intermediate energy 
peaked classes \citep[LBLs, HBLs, IBLs,][]{PadovaniGiommi95,Laurent99}. 
All but three MOJAVE BL~Lacs are classified as LBLs
\citep[see][NED]{Nieppola06}. 
The three BL~Lacs, $viz.,$ 0422+004, 1538+149 and 1807+698, are classified as IBLs.  
There are no HBLs in the MOJAVE sample. 

The compact flux density selection criteria of the MOJAVE sample biases it 
heavily toward highly beamed blazars with high Lorentz factors 
and small viewing angles.  
While relativistic boosting effects are likely to dominate the source 
characteristics, the extensive multi-epoch and multi-wavelength data 
available for the MOJAVE sample on both parsec- and kiloparsec-scales, 
provides us with unique constraints to test the unified scheme.
Throughout the paper, we adopt the cosmology in which
$H_0$=71 km s$^{-1}$ Mpc$^{-1}$, $\Omega_m$=0.27 and $\Omega_{\Lambda}$=0.73.
Spectral index, $\alpha$, is defined such that, flux density
$S_\nu$ at frequency $\nu$, is $S_\nu\propto\nu^{-\alpha}$.

\section{DATA REDUCTION AND ANALYSIS}
We observed seven MOJAVE quasars at 1.46 GHz with the Very Large Array 
\citep[VLA,][]{Napier83} in the A-array configuration (typical synthesized beam 
$\sim1.5\arcsec$) on June 30, 2007 (Program ID:AC874). 
Sixty MOJAVE sources were previously observed by us with the VLA A-array
and presented in \citet{Cooper07}. Data for the remaining sources were
reduced directly using archival VLA A-array data, or obtained 
through published papers. The data reduction was carried out 
following standard calibration and reduction 
procedures in the Astronomical Image Processing System (AIPS).

\begin{figure}[ht]
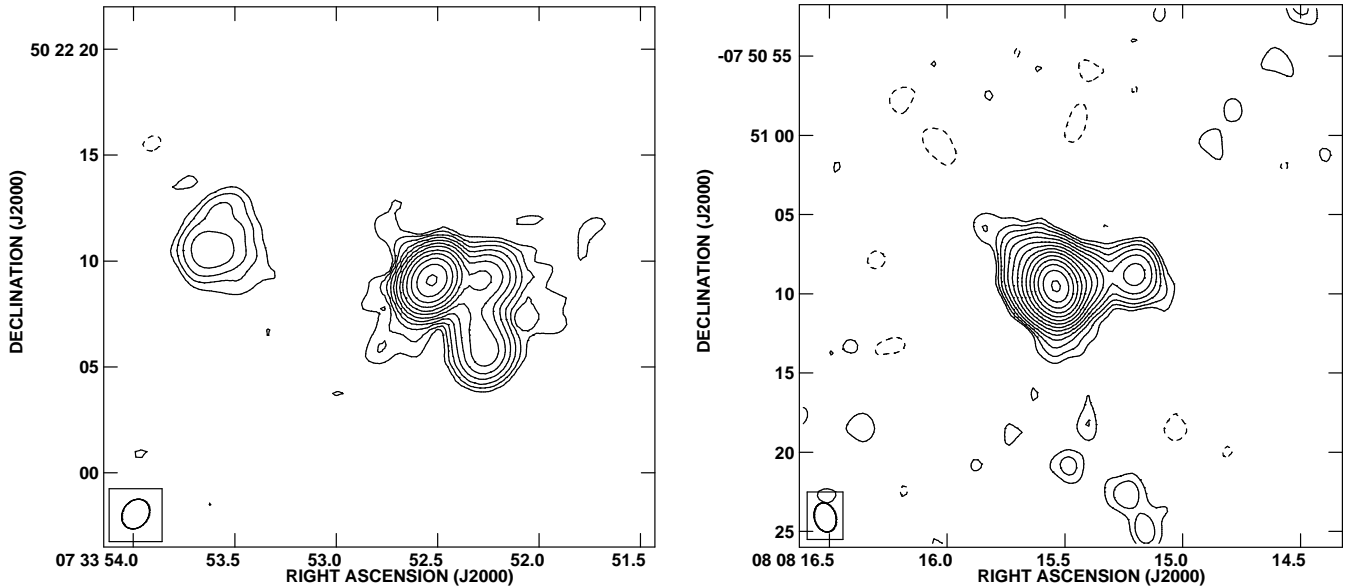

\centerline{
\includegraphics[width=9cm]{figure1a.ps}
\includegraphics[width=9cm]{figure1b.ps}}
\caption{\small 1.4~GHz VLA images of the quasars 0730+504 (Left) 
and 0805$-$077 (Right).
The contours are in percentage of the peak surface brightness and increase 
in steps of 2. The lowest contour levels and peak surface brightness are
(Left) $\pm$0.042, 672 mJy~beam$^{-1}$ and 
(Right) $\pm$ 0.010, 1.43 Jy~beam$^{-1}$.} 
\label{fig:0733}
\end{figure}

\begin{figure}[ht]
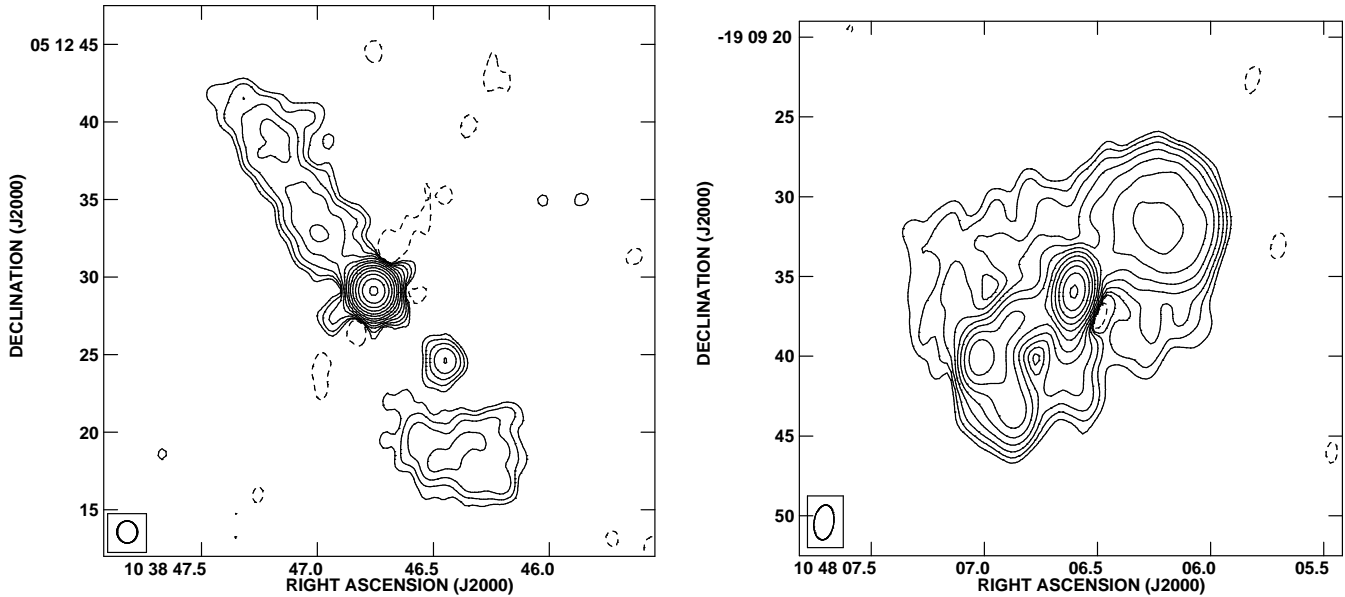

\centerline{
\includegraphics[width=9cm]{figure2a.ps}
\includegraphics[width=9cm]{figure2b.ps}}
\caption{\small 1.4~GHz VLA images of 1036+054 (Left) and 1045$-$188 (Right).
The contours are in percentage of the peak surface brightness and increase 
in steps of 2. The lowest contour levels and peak surface brightness are
(Left) $\pm$0.021, 892 mJy~beam$^{-1}$ and 
(Right) $\pm$ 0.042, 724 mJy~beam$^{-1}$.}
\label{fig:1038}
\end{figure}

\begin{figure}[ht]
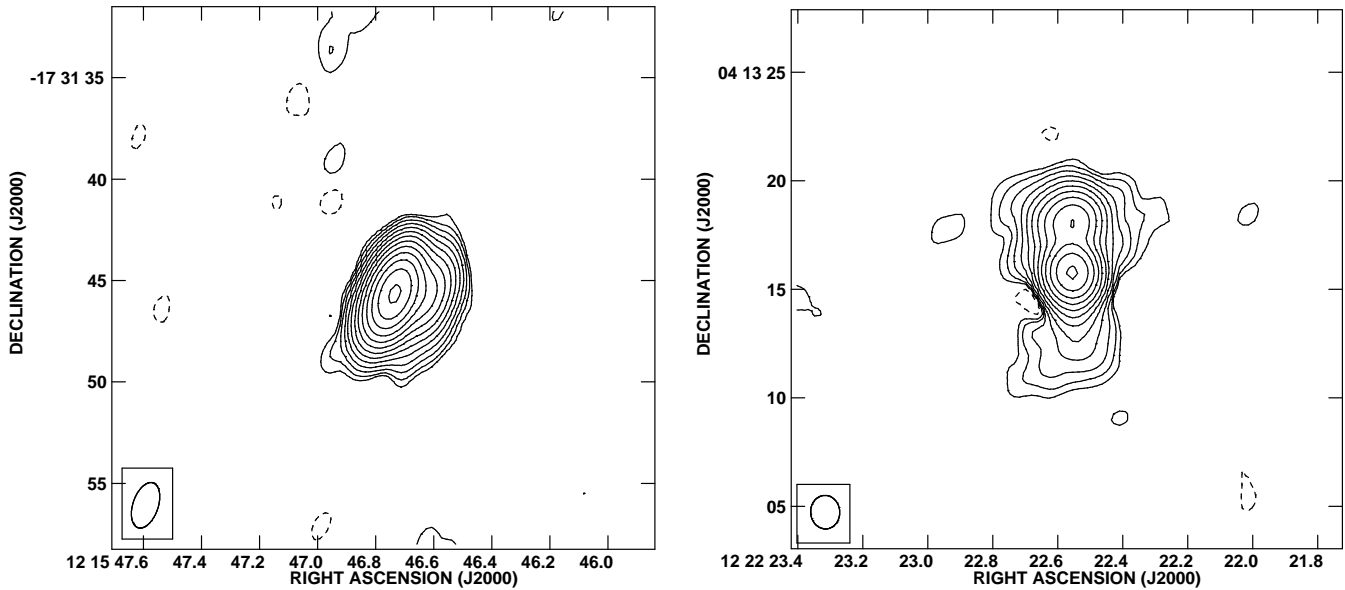

\centerline{
\includegraphics[width=9cm]{figure3a.ps}
\includegraphics[width=9cm]{figure3b.ps}}
\caption{\small 1.4~GHz VLA images of 1213$-$172 (Left) and 1219+044 (Right).
The contours are in percentage of the peak surface brightness and increase 
in steps of 2. The lowest contour levels and peak surface brightness are
(Left) $\pm$0.021, 1.66 Jy~beam$^{-1}$ and 
(Right) $\pm$ 0.042, 549 mJy~beam$^{-1}$.}
\label{fig:1215}
\end{figure}

Our new observations included some new expanded VLA (EVLA) antennas and the 
observations were made in a spectral line (4-channel pseudo-continuum) mode, 
which resulted 
in a total effective bandwidth of ($2\times21.875=$) 43.75~MHz. Four $\approx9$ 
minute scans of each source were interspersed with 1 minute scans of a 
suitable phase calibrator. 0713+438 was used as the bandpass 
calibrator for the experiment. 
After the initial amplitude and phase calibration, AIPS tasks CALIB and 
IMAGR were used iteratively to self-calibrate \citep{Schwab80} 
and image the sources.
The resultant {\it rms} noise in the maps was typically of the
order of 0.06~mJy~beam$^{-1}$.
The quasar 1124$-$186 was observed with an incorrect source position, which 
resulted in significant beam smearing. We instead obtained archival VLA data 
for this source.
The new radio images of the six sources are presented in Figures 
\ref{fig:0733}, \ref{fig:1038}, and \ref{fig:1215}. 

Since most of the MOJAVE sources have highly compact radio structures,
most of the archival data comprised of snapshot observations 
where the sample sources were observed as phase calibrators. 
By choosing archival datasets with exposure times $\gtrsim$10 minutes, we were 
able to obtain maps with typical {\it rms} noise levels of 
$\sim0.15$~mJy~beam$^{-1}$.
We present previously unpublished images of 21 blazars in Appendix A.

A compilation of the basic parameters for each source is given in 
Table~\ref{tabsample}. The integrated flux densities for the cores were 
obtained in AIPS using the Gaussian-fitting task JMFIT, while the total 
flux densities were obtained by putting a box around the source, using the 
AIPS verbs TVWINDOW and IMSTAT.
{The extended radio flux density was obtained by subtracting the core flux 
density from the total radio flux density.

All the maps were created with uniform weighting using a ROBUST parameter 
of 0 in the AIPS task IMAGR. We also obtained core and lobe flux densities
from maps with different weighting schemes (by using ROBUST parameters
$-$5 and +5, respectively)} {for a fraction of the sample}. We found 
that the integrated core flux density estimates differed typically
by less than 1\%, between different weighting schemes.
The extended flux density values differed typically by $<2\%$, with
a large majority of sources showing less than a 5\% difference.
Only in a handful of sources, where there seemed to be an unresolved
compact component close to the core (e.g., 1038+064, 1417+385), 
was the difference between different weighting schemes significant 
(10\%$-$20\%). However, following our radio galaxy study \citep{Kharb08a},
we have found that nearly 15\%$-$20\% of the extended flux density could be 
getting lost in A-array observations, compared to combined-array observations.
Therefore, we conclude that the lack of short spacings in the A-array 
observations is far more detrimental to the determination of accurate
extended flux values than (not) adopting different weighting schemes on the 
A-array data to obtain core and extended flux densities. 
The weighting scheme approach does suggest that 
the errors in the extended flux density values  
are typically of the order of 2\%$-$5\%, but could be of the order
of 10\%$-$20\% for sources where a compact component close to the core
is not clearly resolved. 

\section{RESULTS}
\subsection{Extended Radio Power}
\begin{figure}[ht]
\centerline{
\includegraphics[width=9cm]{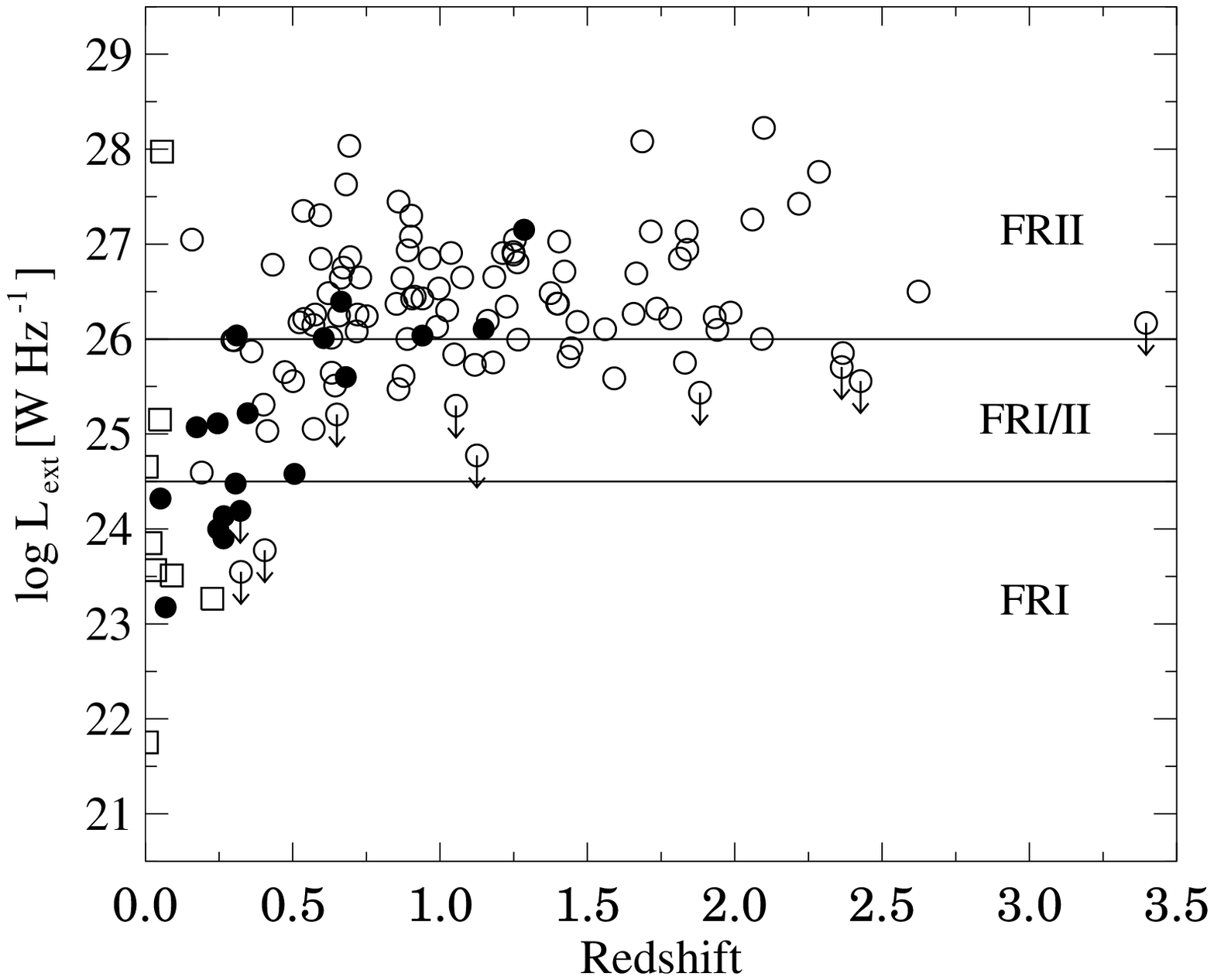}
\includegraphics[width=9cm]{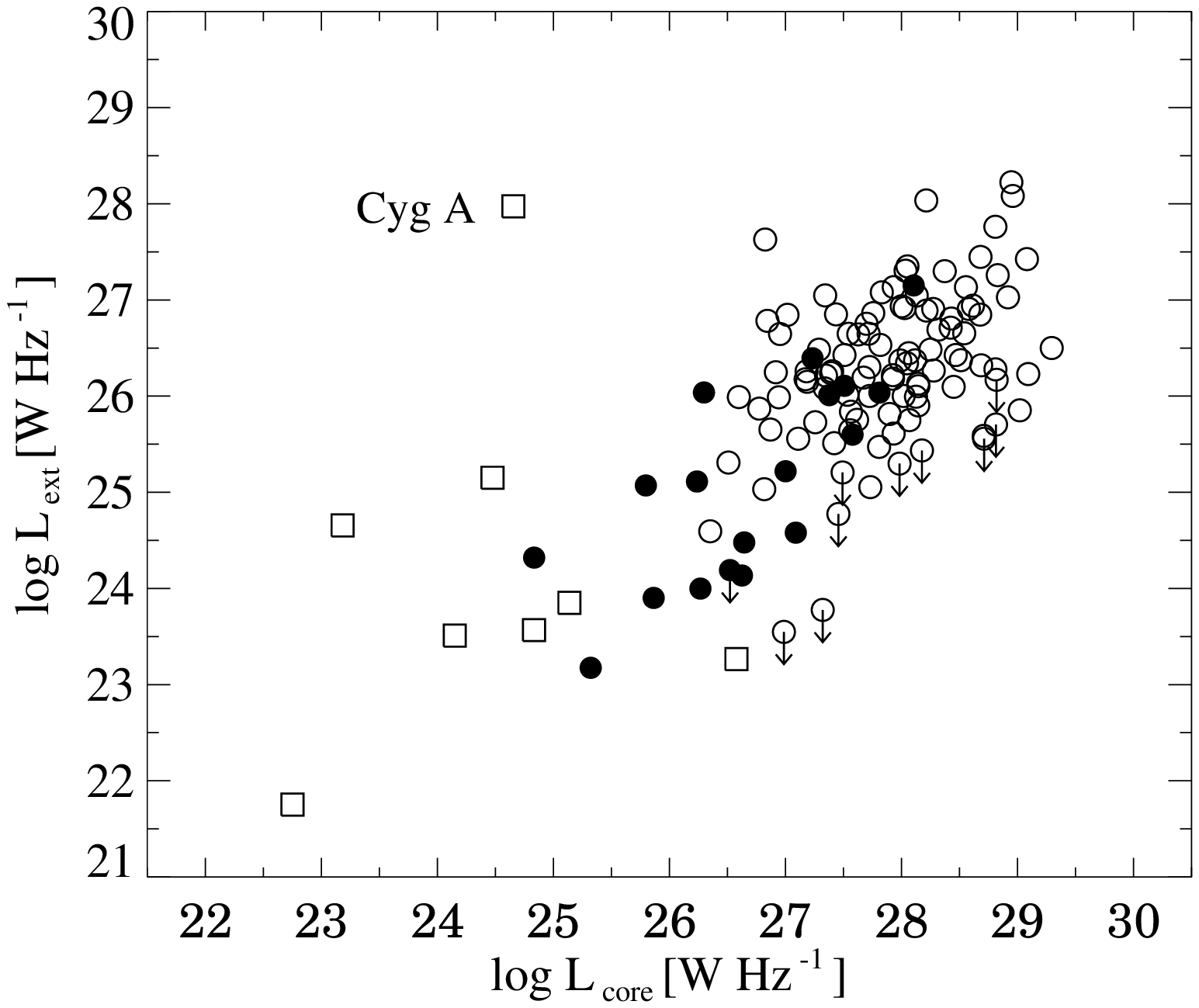}}
\caption{\small (Left) 1.4~GHz extended luminosity versus redshift. 
(Right) 1.4~GHz core versus extended luminosity. 
Open and filled 
circles denote quasars and BL~Lacs, respectively, while open squares denote 
radio galaxies. Core-only sources are represented as upper limits.}
\label{fig:z}
\end{figure}

The 1.4~GHz extended radio luminosities for the MOJAVE sources are 
plotted against redshift in Figure~\ref{fig:z}. 
Sources with no discernible extended 
emission are represented as upper limits.
The solid lines indicate the FRI$-$FRII divide (extrapolated from 178~MHz 
to 1.4 GHz assuming a spectral index, $\alpha= 0.8$), following 
\citet{LedlowOwen96} and \citet{Landt06}. The right hand panel of 
Figure~\ref{fig:z} demonstrates the close relation between the radio core 
and extended luminosity (Table~\ref{tabcorrel}).
Using the partial correlation regression analysis routine in
IDL (P$\_$CORRELATE), we found that the linear correlation
between log$L_{core}$ and log$L_{ext}$ is strong even with the effects of
luminosity distance (log$D_{L}$) removed (partial correlation coefficient, 
$r_{XY.Z}$, = 0.303, $t$ statistic = 3.41, two-tailed probability that the variables
are not correlated{\footnote{Calculated using the VassarStats statistical
computation website, 
http://faculty.vassar.edu/lowry/VassarStats.html}}, $p$ = 0.0009).
We note that the 18 BL Lacs (ones with redshift information) alone fail to show
a correlation between log$L_{core}$ and log$L_{ext}$, when the effects of
luminosity distance are removed. The implication of this finding 
is discussed ahead in \S3.3.

The salient features of the extended radio emission in the MOJAVE blazars
are:
\begin{enumerate}
\item[(i)] Six of the 22 BL Lacs ($\sim27\%$) have FRII radio powers 
(log$L_{ext}^{1.4} > 26$). Four of these 
($viz.,$ 0235+164, 0716+714, 0808+019, 2131$-$021) also have hot spots 
like quasars. 
\item[(ii)] Five BL Lacs ($\sim23\%$) fall in the FRI/II power range  
($24.5 <$ log$L_{ext}^{1.4} < 26$). Two of these ($viz.,$ 0823+033, 1803+784)  
also appear to have hot spots.
\item[(iii)] Seven BL Lacs have extended luminosity 
log$L_{ext}^{1.4} < 24.5$ and could be regarded as the true beamed counterparts
of FRI radio galaxies. 
However, three of these ($viz.,$ 0754+100, 0851+202, 1807+698) also appear to exhibit 
hot~spot-like features. Hot spots may also be present in all four BL~Lacs with
no redshift information. Overall, nearly 60\% of the MOJAVE BL~Lacs 
appear to have hot spots.
{The hot spots in some BL~Lacs are however not as 
bright or compact as those observed in quasars.}
\item[(iv)] Excluding the upper limits, 22 quasars ($\sim22\%$) fall in the FRI/II power range.
\item[(v)] 10 sources ($\sim7\%$, 9 quasars, 1 BL Lac) do not show any 
extended emission $-$ these are discussed in \S3.4.
\end{enumerate}

Based on the extended emission at 1.4~GHz, we can conclude that 
a substantial fraction of the MOJAVE BL Lacs have both radio powers and 
radio morphologies like FRIIs or quasars. A substantial fraction of the
MOJAVE quasars lie in the intermediate (FRI/II) luminosity range.
These results are consistent with a number of previous radio studies.
Using a large sample of radio core-dominated AGNs, \citet{Murphy93} observed that 
many high redshift ($z>0.5$) BL~Lacs lie above the FRI/FRII luminosity 
division. Using a sample of 17 BL~Lacs, mostly belonging to the 1-Jy sample,
\citet{Kollgaard92} suggested that about 12\% $-$ 30\% of BL Lacs 
in a radio flux-limited sample may be bona fide FRIIs. 
Based on extended radio emission and strong emission lines at one or more
epochs, \citet{Rector01} concluded that many radio-selected BL~Lacs 
belonging to the 1 Jy sample \citep{Stickel91} cannot be beamed FRIs, but
are more likely to be beamed FRIIs. 
\citet{Cara08} found similar intrinsic parent luminosity functions for 
the MOJAVE sample (consistent with FRIIs) irrespective of whether the 
BL~Lacs were included or not.

\subsection{Radio Morphology}
We observe from Figures \ref{fig:0733}--\ref{fig:1215} that
all but two sources show two-sided radio structures. The variety of radio 
structures observed here are reasonably representative of the entire sample 
\citep[e.g., see][]{Cooper07}, as well as those 
previously observed in other quasar surveys 
\citep[e.g.,][]{Gower84,Antonucci85,PearsonReadhead88,Murphy93}. 
The quasars often show two-sided radio structures with hot spots on
one or both sides of the core. However, one-sided morphologies with
no discernable compact or diffuse emission on the other side of the core, are 
also common. 

While determining if the blazars had an FRII type radio morphology,
we relied initially on the presence of one or more compact hot 
spots at the leading edge of the radio lobes. When no clear hot spots 
were visible on either side of the cores, we resorted to the traditional
``edge-brightened'' definition for FRII sources, $i.e.,$ when the brightest
radio emission was furthest from the core, we classified the source as an FRII.
The remaining sources with no clear hot spots, and the brightest extended emission
closest to the cores, were classified as FRI types. Even then, it was sometimes
difficult to consign the sources to FRI or FRII classes,
and we have listed more than one morphology for some sources in
Table~\ref{tabsample}.

Many of the MOJAVE sources show distinctly curved kiloparsec-scale jets. 
Straight jets are a rarity in the sample. Many 
MOJAVE quasars exhibit large parsec-to-kiloparsec jet misalignments. 
While any intrinsic curvature in the jets is likely to be highly exaggerated 
by projection effects in these low-viewing angle blazars, many sources seem 
intrinsically distorted.
We discuss in \S4.1 how the MOJAVE selection criteria might be preferentially
picking up bent jets. 
Apart from highly curved jets, many sources show distinct hot spot-like 
features both closer to the core and at the jet termination points, similar to
the wide angle tail (WAT) radio galaxies \citep[e.g.,][]{O'Donoghue90}.

Four MOJAVE blazars have not yet been identified as quasars or BL~Lacs 
(Table~\ref{tabsample}). A new VLA image of one of the unidentified sources 
($viz.,$ 1213$-$172) is presented in Figure~\ref{fig:1215}. Another unidentified 
source (0648$-$165) has a similar core-dominant radio morphology \citep{Cooper07}.
A third source (2021+317) has a distinct core-halo morphology resembling a 
BL Lac object (see Appendix A), while the fourth (0446+112) has a straight jet with 
a possible hot spot at the end \citep{Cooper07}.

\subsection{Parsec-scale Jet Speeds and Extended Luminosity}
\begin{figure}[ht]
\centerline{
\includegraphics[width=9cm]{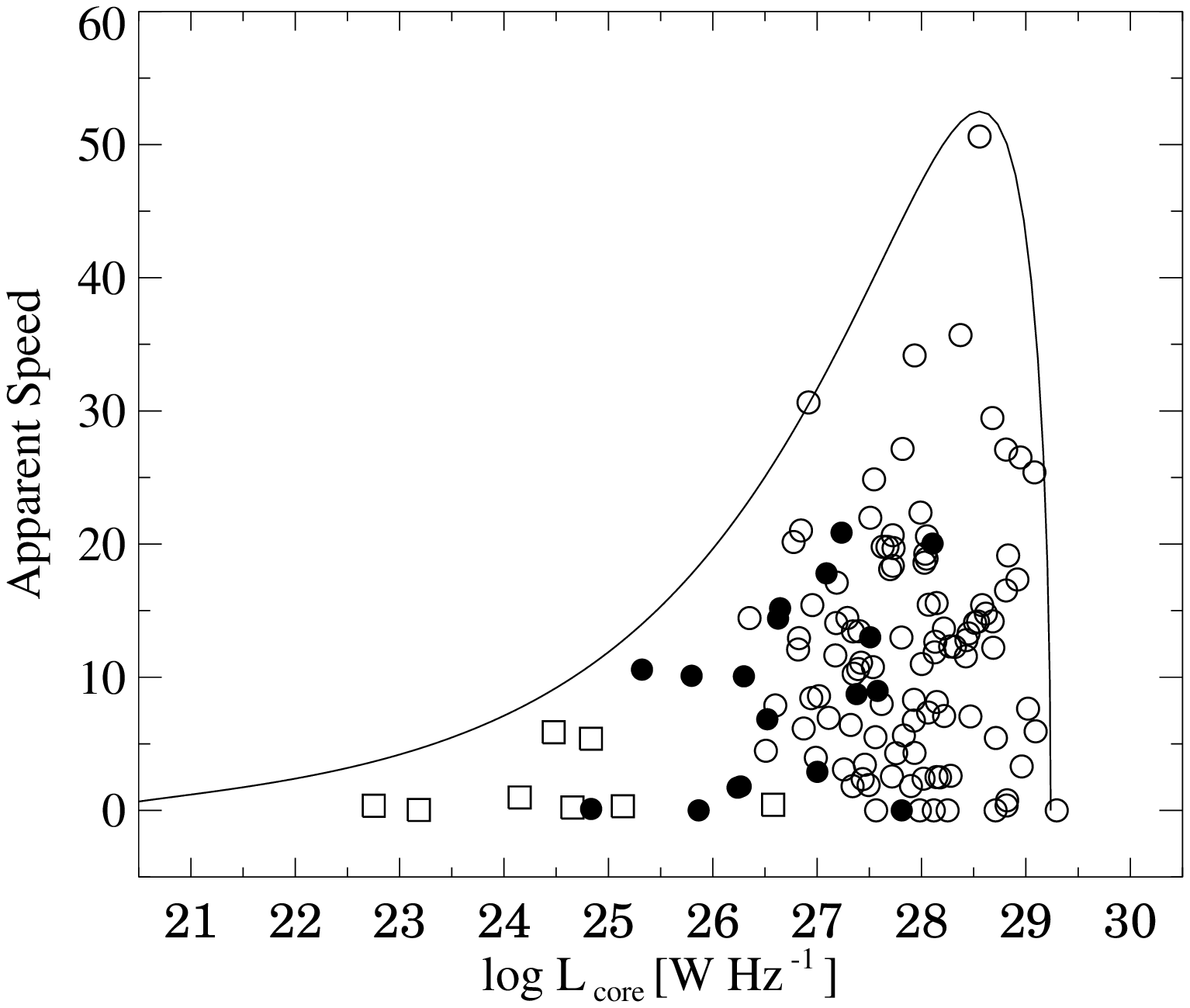}
\includegraphics[width=9cm]{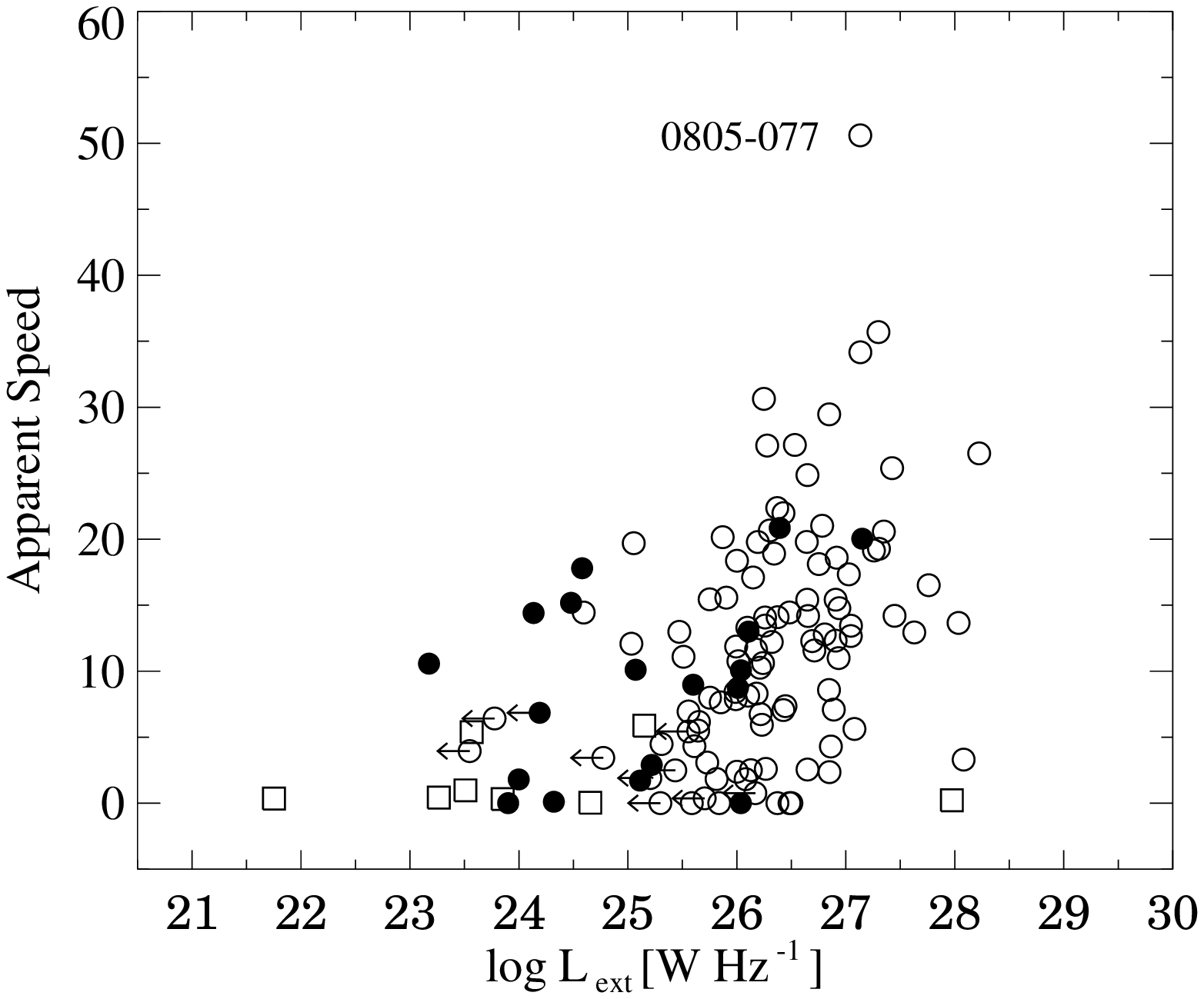}}
\caption{\small (Left) 1.4~GHz core luminosity versus the apparent jet speed.
The aspect curve assumes $\gamma=52$, $L_{int}=5\times10^{24}$ and $p\approx$2.
(Right) 1.4 GHz extended luminosity versus the apparent jet speed.
Open and filled 
circles denote quasars and BL~Lacs, respectively, while open squares denote 
radio galaxies. Core-only sources are represented as upper limits.}
\label{fig:beta}
\end{figure}

The MOJAVE and 2~cm survey programs have provided apparent parsec-scale
jet speeds ($\beta_{app}$) for nearly the entire MOJAVE sample, by
monitoring structural changes over timescales of approximately a decade.
These have been previously tabulated by \citet{Kellermann04}, and most
recently by \citet{Lister09a}. For our analysis below, we use a 
$\beta_{app}$ value that represents the fastest moving feature in 
that source.

The parsec-scale apparent jet speeds and kiloparsec-scale radio core luminosity 
(Fig.~\ref{fig:beta}) are related by standard beaming relations. The aspect 
curve in Figure~\ref{fig:beta} assumes: $L=L_{int}\times\delta^p$,
$\beta_{app}=\beta$~sin~$\theta/(1-\beta$~cos~$\theta)$, where the
Doppler factor, $\delta=1/(\gamma~(1-\beta$~cos~$\theta))$ 
and $\beta=v/c$. The best-fit values for the curve are 
$\gamma=52$, $L_{int}=5\times10^{24}$ and $p\approx2$. 
The relation between parsec-scale apparent jet speeds and parsec-scale
core luminosity for the MOJAVE sample is discussed in greater detail by
\citet{Cohen07}, and \citet{Lister09a}.

We find that the parsec-scale apparent jet speeds are correlated significantly
with the extended radio luminosity (Fig.~\ref{fig:beta}, Table~\ref{tabcorrel}).
Partial regression analysis shows that the linear correlation between
log$L_{ext}$ and $\beta_{app}$ with the effects of luminosity distance removed,
is still highly significant ($r_{XY.Z}$ = 0.273, $t$ statistic = 3.05,
two-tailed probability, $p$ = 0.0028).
This implies that more radio powerful sources have faster radio jets.
Since extended radio luminosity is correlated with the core 
luminosity (Fig.~\ref{fig:z}, Table~\ref{tabcorrel}), and the core luminosity 
is related to the apparent jet speed, we tested the correlation with a 
partial correlation test between extended radio luminosity and apparent jet speeds, 
after removing the effects of radio core luminosity.
This too yielded a statistically significant 
correlation between extended radio luminosity and parsec-scale apparent jet 
speeds ($r_{XY.Z}$ = 0.246, $t$ statistic = 2.72, $p$ = 0.0075).

The significant implication of this result is that faster jets are 
launched in AGNs with larger kiloparsec-scale lobe luminosities. It would 
therefore appear that the fate of the AGN is decided at its conception.
This result undermines the role of the kiloparsec environment on the radio
power of a given source. It also indicates that the 1.4~GHz extended emission 
is indeed related to jet kinetic power, contrary to some 
suggestions in the literature \citep[e.g.,][]{Birzan04}. Given that there is a 
large overlap in radio powers between quasars and BL~Lac objects, it can be 
concluded that most quasars have faster jets than most BL~Lac objects. 

The second noteworthy inference that can be drawn from Figure~\ref{fig:beta}
is that there is a continuous distribution of parsec-scale jet speeds going 
from BL~Lacs to quasars. This supports the idea that the distinction 
between the BL~Lac and quasar population, at an equivalent width of 
5$\AA$, is essentially an arbitrary one at least in terms of radio jet 
properties \citep[see also][]{ScarpaFalomo97}. 

However, we note that inspite of the two blazar classes displaying
a smooth continuation in properties (as is evident in Fig.~\ref{fig:beta} and
other plots), they sometimes reveal differences in the correlation test results
when considered separately. In Table~\ref{tabcorrel} we have listed
the correlation test results separately for quasars and BL Lacs, if
they differed from results for the combined blazar population.
BL~Lacs considered alone sometimes failed to exhibit the correlations
observed in quasars (also see \S3.1). 
Apart from the possible effects of small number statistics, a
simple interpretation of this finding could be
that unlike the quasars, the BL~Lacs do not 
constitute a homogeneous population. As has been
previously suggested in the literature \citep[e.g.,][]{Owen96,Cara08,Landt08}, 
the BL Lacs
might constitute the beamed population of both FRI and FRII radio galaxies. 
We note that \citet{Giroletti04} have demonstrated that HBLs (which are
absent in MOJAVE) conform to the standard unification scheme with respect to FRI 
radio galaxies. This again supports the proposed inhomogenity within the
BL Lac class.  

Preliminary Monte Carlo simulations of a population of AGN similar to that 
of MOJAVE also show a correlation between apparent jet speed and
extended luminosity (Cooper et al., in preparation). 
The simulations are based on the luminosity function of \citet{PadovaniUrry92},
and assume that the extended luminosity is unbeamed and proportional 
to the intrinsic unbeamed parsec-scale luminosity. We are currently 
incorporating the luminosity function derived from the MOJAVE sample by 
\citet{Cara08} into the simulations.

\subsection{Parsec-to-kiloparsec Jet Misalignment}

\begin{figure}[ht]
\centerline{
\includegraphics[width=9cm]{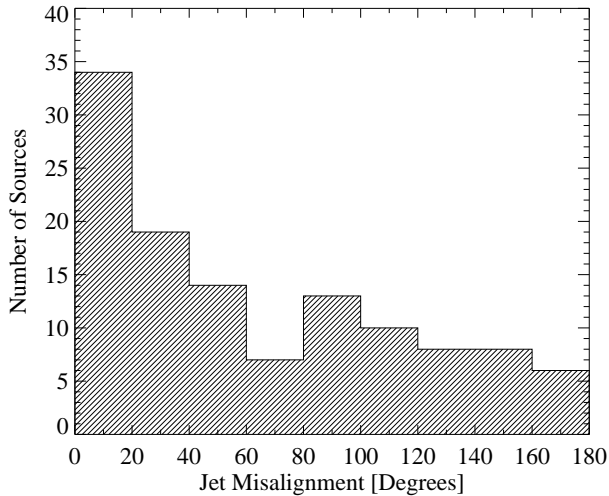}}
\caption{\small Misalignment angle between the parsec-scale and the 
kiloparsec-scale jet for the MOJAVE sources. 
There are signatures of a weak $90\degr$ bump.}
\label{fig:Delta}
\end{figure}

Apparent misalignment in the jet direction from parsec to kiloparsec scales
is commonly observed in blazars \citep[e.g.,][]{PearsonReadhead88}. 
This misalignment could either be due to actual large bends in the jets, 
or due to small bends that are amplified by projection.
We present the parsec- to kiloparsec-scale jet misalignment for the MOJAVE sources
in Figure~\ref{fig:Delta}. 
The procedure for estimating the parsec-scale jet position angles (PAs) 
is described in \citet{Lister09}. Kiloparsec-scale jet position angles were 
determined for the FRII sources using the brightest hot spots, especially when 
no jet was clearly visible,
under the assumption that the brightest hot spot indicates the approaching
jet direction due to Doppler boosting. 
The AIPS procedure TVMAXFIT was used to obtain accurate peak pixel positions
of the core and hot spot, which were then used in the AIPS verb IMDIST 
to obtain the jet position angle.
For the FRI-type sources we used TVMAXFIT for the core pixel position, and 
the AIPS verb CURVALUE to get the pixel position roughly towards the center 
and end of the broad jet/lobe. Finally IMDIST was used to obtain the jet 
position angle. 
Note that unlike regular radio galaxies, FRI blazars are typically one-sided,
which makes the approaching jet direction easy to identify.
Due to the wide jet/lobes in FRIs, the kiloparsec-scale jet 
position angle could be uncertain by 10$\degr$ to 15$\degr$.
When a jet feature appeared only partially resolved ($viz.,$ 1038+064, 
1417+385), a two Gaussian component model was used in JMFIT to obtain pixel 
positions of the core and jet feature.

We note that our approach ($viz.,$ of using the brightest hot spot to
determine the approaching jet direction, when no jet was clearly visible)
differs from that adopted by some 
other authors \citep[e.g.,][]{Xu94} who assume that, when no jet is visible, the 
hot spot on the side of the parsec-scale jet, represents the kiloparsec-scale 
jet direction. Our approach has led us to identify many 
more sources ($\sim30\%$) with 
misalignment angles greater than 90$\degr$.

For an otherwise straight jet having a single bend,
the apparent bending angle ($\eta$) is related to the intrinsic
bending angle ($\psi$) through the relation,
$\cot~\eta=\frac{\cot~\psi~\sin~\theta - \cos~\theta~\cos~\phi}{\sin~\phi}$,
where $\theta$ is the angle to line of sight, and $\phi$ is the azimuth of 
the bend \citep{Appl96}.
The fact that we see a majority of sources having close to zero apparent 
misalignment implies one of the following scenarios:
(i) the intrinsic misalignment angle is small;
(ii) both the intrinsic misalignment and jet inclination angles are large;
(iii) the azimuth of the bend is close to 180$\degr$, $i.e.,$ the plane of the
bend is perpendicular to the sky plane. Since scenario (ii) is unlikely for 
these blazars, possibilities (i) and (iii) are more favorable.
In \S4, we explain how the MOJAVE selection criteria could make it 
biased towards bent jets. 

\citet{PearsonReadhead88} and \citet{ConwayMurphy93} reported an unexpected 
secondary peak at $\Delta$PA = $90\degr$ in the misalignment angle 
distribution. The signatures of a weak $90\degr$ bump appear to be
present in Figure~\ref{fig:Delta}.
\citet{ConwayMurphy93} concluded that while the largely 
aligned population could be explained by straight parsec-scale jets and small
intrinsic bends between parsec- and kiloparsec-scales, the secondary peak sources
must have gently curving (low-pitch) helical parsec-scale jets.
This helical distortion could arise due to Kelvin-Hemlholtz instabilities 
or precession of the jet ejection axes. 
We reach a similar conclusion on the jet misalignments, following a
different approach in \S4.3.

\section{DISCUSSION}
\subsection{Selection Effects in MOJAVE}
The MOJAVE survey appears to select many sources with intermediate radio powers
and radio morphology (e.g., high radio power BL~Lacs, some with hot spots, and 
low radio power quasars). This could be a result of the MOJAVE selection criteria, 
which are based on the relativistically-boosted, high radio frequency (15~GHz) 
parsec-scale flux densities.
This would make it more likely to encompass a much larger range in 
intrinsic radio powers, compared to other samples selected on the basis of 
total kiloparsec-scale flux-densities (e.g., the 3CR sample). 
The intrinsic radio luminosity function derived for the 
MOJAVE sample by \citet{Cara08} also supports this view. 

Furthermore, as the MOJAVE survey picks up bright radio cores, most of the
sources could either be fast jets pointed towards us, or curved/bent
jets that have at least some portion of the jet aligned directly into
our line of sight, at at least one epoch \citep[e.g.,][]{Alberdi93}. 
The MOJAVE survey could therefore be biased towards bent jets.
Note that in the wide angle tail quasars, the hot spots closer to the cores could 
be produced at the base of the plumes \citep[e.g.,][]{HardcastleSakelliou04}, or 
indicate a sharp bend in the jet \citep[e.g.,][]{Alberdi93,Jetha06}.
We discuss the possibility of beamed WAT quasars in the MOJAVE sample further 
in \S4.3.

\subsection{Orientation Effects in the Sample}
Since the MOJAVE sample consists almost entirely of blazars, the sources are 
expected to have jets aligned close to our line of sight. Our
examination of orientation effects in the sample, however, using two different
statistical orientation indicators, $R_c$ and $R_v$, reveals 
that a range of orientations must in fact be present in order to
account for several observed trends. $R_c$ and $R_v$ also reveal, to some extent,
dissimilar trends. We finally reach the conclusion that $R_v$ is a better 
indicator of orientation for this blazar sample.

\begin{figure}[ht]
\centerline{
\includegraphics[width=9cm]{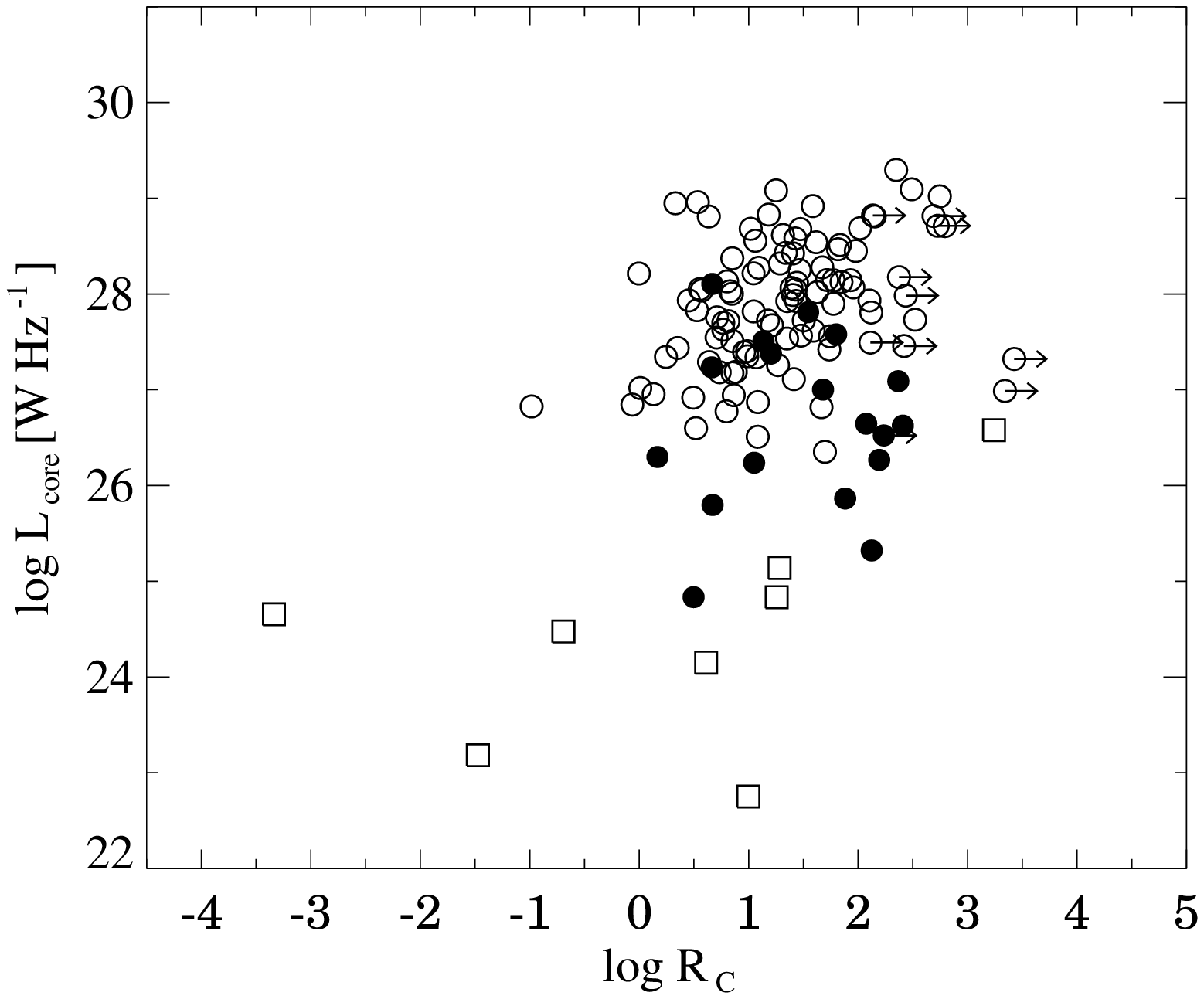}
\includegraphics[width=9cm]{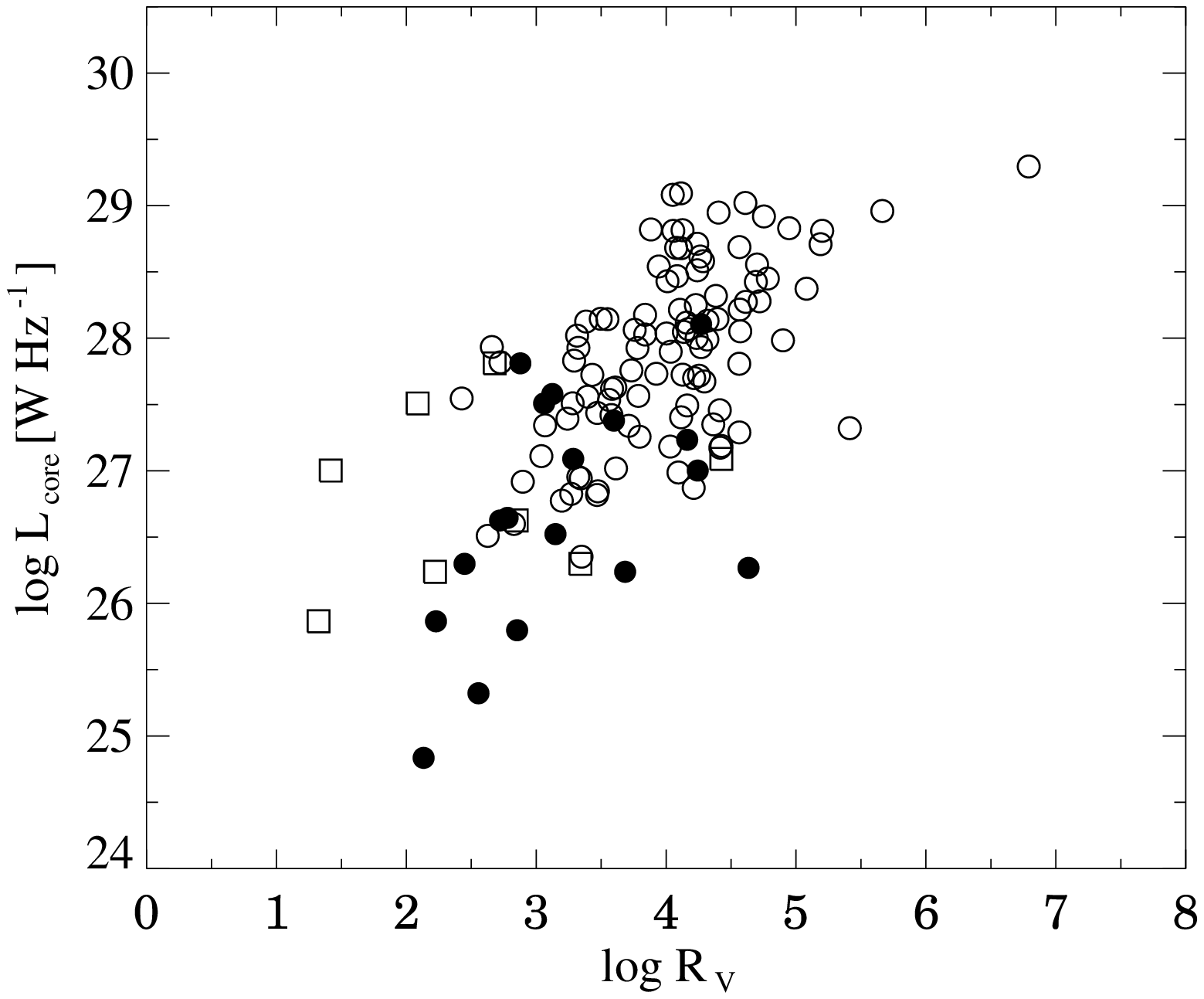}}
\caption{\small The orientation indicators, $R_c$ (Left) and $R_v$ (Right) 
versus the 1.4 GHz core luminosity.
Open and filled 
circles denote quasars and BL~Lacs, respectively, while open squares denote 
radio galaxies. Core-only sources are represented as upper limits.}
\label{fig:RcCore}
\end{figure}

\begin{figure}[ht]
\centerline{
\includegraphics[width=9cm]{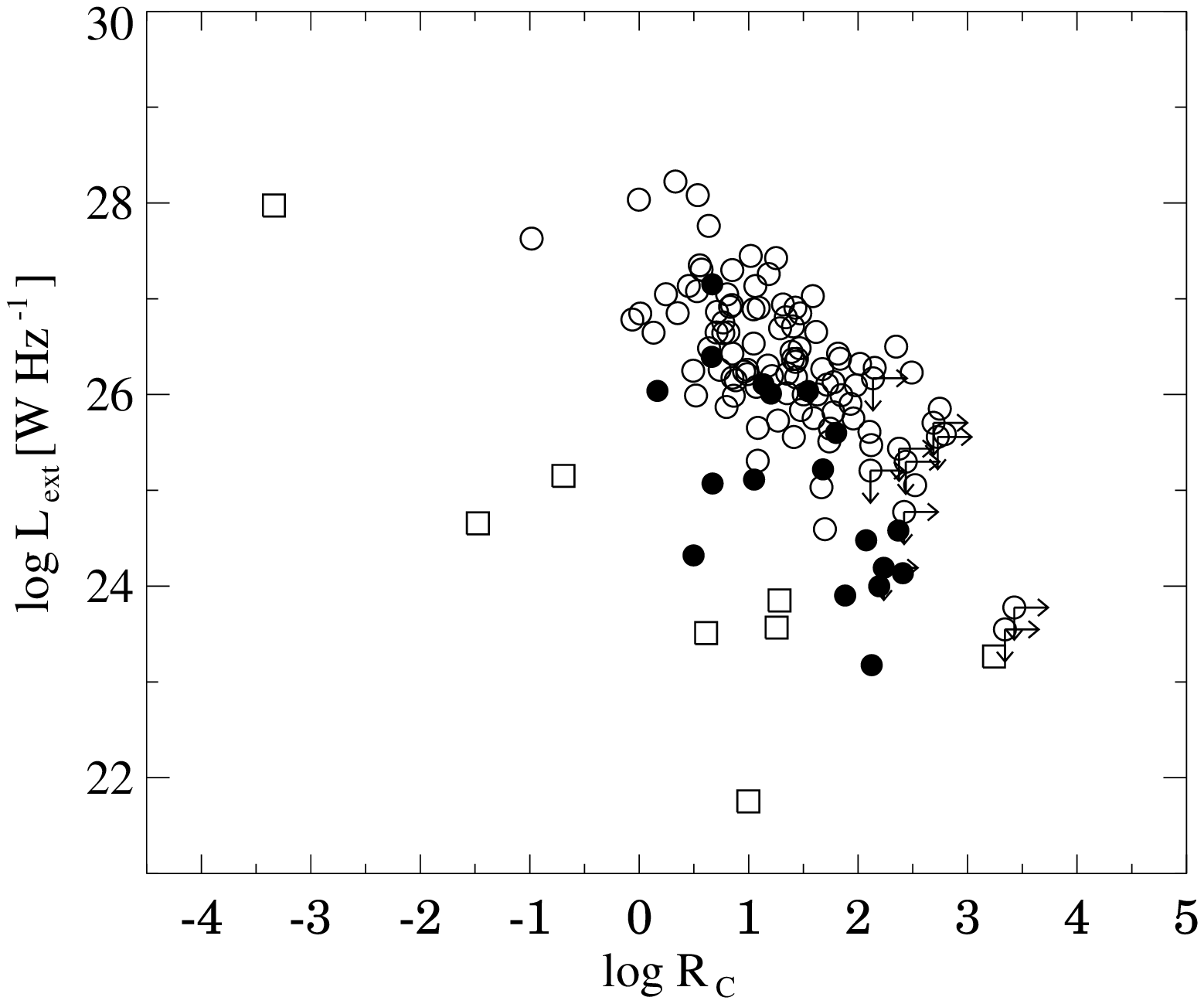}
\includegraphics[width=9cm]{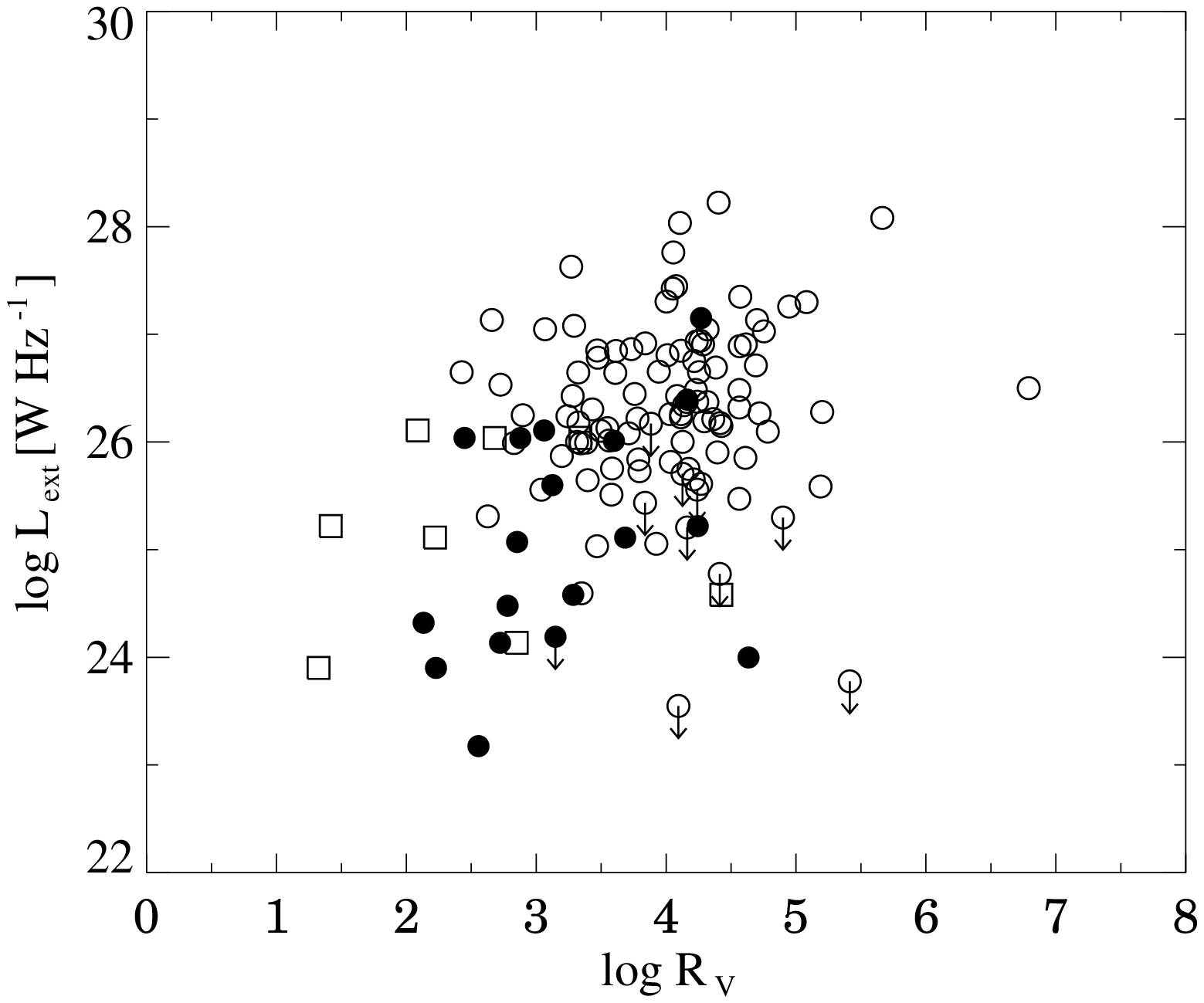}}
\caption{\small The orientation indicators, $R_c$ (Left) and $R_v$ (Right) versus 
the 1.4 GHz extended luminosity.
Open and filled 
circles denote quasars and BL~Lacs, respectively, while open squares denote 
radio galaxies. Core-only sources are represented as upper limits.}
\label{fig:RcExt}
\end{figure}

The ratio of the beamed radio core flux density ($S_{core}$) to the unbeamed 
extended radio flux density ($S_{ext}$), $viz.,$ the radio core prominence parameter 
($R_c$), has routinely been used as a statistical 
indicator of Doppler beaming and thereby orientation 
\citep{OrrBrowne82,KapahiSaikia82,Kharb04}. The $k$-corrected $R_c$
($=\frac{S_{core}}{S_{ext}}(1+z)^{\alpha_{core} - \alpha_{ext}}$, with 
$\alpha_{core}=0$, $\alpha_{ext}=0.8$) is plotted against the 
radio core luminosity in 
Figure~\ref{fig:RcCore}. As expected, $R_c$ is correlated with the radio 
core luminosity (Table~\ref{tabcorrel}). $R_c$ however, shows a 
significant anti-correlation with respect to the extended radio luminosity 
(Fig.~\ref{fig:RcExt}).
$R_c$ does not show a correlation with jet misalignment (Table~\ref{tabcorrel}). 
In fact the quasars considered alone showed
a weak anti-correlation between $R_c$ and jet misalignment (implying
that the more core-dominant sources have smaller jet misalignments).
These results are unexpected, because 
in the former case, the extended radio luminosity is expected to be
largely unbeamed, while in the latter, the jet misalignment is affected by
projection and therefore orientation
\citep[in the sense that the more core-dominant sources
show larger jet misalignments, e.g.,][]{KapahiSaikia82}. 
As an aside, \citet{Brotherton96} have noted that the anti-correlation
between log$R_c -$ log$L_{ext}$ may be suggesting that core-dominated quasars are 
intrinsically fainter than their lobe-dominated counterparts. This would 
be consistent with the idea that the MOJAVE sources span a large range in 
intrinsic radio powers ($cf.$ \S4.1).

We also examined the relationship between $R_c$ and apparent jet speed.
While both $R_c$ and $\beta_{app}$ depend on the Lorentz factor and orientation, 
many of the viewing angles may be inside the critical angle for maximum
superluminal speed, which would spoil any correlation. However, the quasars 
considered alone do show a significant anti-correlation, contrary to 
expectations. But as we see below, the alternate orientation indicator, $R_v$, 
does show the expected behavior.

\citet{WillsBrotherton95} defined $R_v$ as the ratio of the radio core 
luminosity to the $k$-corrected absolute V-band magnitude ($M_{abs}$): 
$log R_v = log \frac{L_{core}}{L_{opt}} = (log L_{core}+ M_{abs}/2.5) - 13.7$,
where $M_{abs}=M_V - k$, and the $k$-correction is, 
$k=-2.5~log~(1+z)^{1-\alpha_{opt}}$ with the optical spectral index,
$\alpha_{opt} = 0.5$. $R_v$ is suggested to be
a better orientation indicator then $R_c$ since the optical luminosity is
likely to be a better measure of intrinsic jet power 
\citep[e.g.,][]{Maraschi08,Ghisellini09} than extended radio
luminosity. This is due to the fact that the optical continuum luminosity is 
correlated with the emission-line luminosity over four orders of magnitude 
\citep{YeeOke78}, and the emission-line luminosity is 
tightly correlated with the total jet kinetic power \citep{RawlingsSaunders91}. 
The extended radio luminosity, on the other hand, is suggested to be 
affected by interaction with the environment on kiloparsec-scales. 

Figure~\ref{fig:RcCore} suggests that $R_v$ is indeed a better indicator of 
orientation as the correlation with radio core luminosity gains in prominence
(see Table~\ref{tabcorrel}). $R_v$ shows a weak positive correlation with 
the extended radio luminosity (Fig.~\ref{fig:RcExt}). 
Partial regression analysis however, shows that the linear correlation
between log$R_v$ and log$L_{ext}$ is no longer significant when the effects of
log$L_{core}$ are removed ($r_{XY.Z}$ = 0.078, $t$ statistic = 0.84,
$p$ = 0.4013). This implies that the extended radio emission
is largely unbeamed.
$R_v$ is correlated with the 
parsec-to-kiloparsec jet misalignment, suggesting that orientation does play 
a role in the observed jet misalignments, at least for the quasars. The lack 
of such a correlation for the BL~Lacs could again be interpreted to be 
consequence of their comprising of not only beamed FRIs (which could have 
intrinsically distorted jets) but beamed FRIIs as well. We discuss this point 
again in \S4.5. Finally, there appears to be no correlation 
between $R_v$ and $\beta_{app}$, as is expected if many blazar viewing angles
are inside the critical angle for maximum superluminal speeds, which would
then result in smaller apparent speeds. 
This result is again consistent with $R_v$ 
being a better indicator of orientation than $R_c$. 

Here we would like to note that the correlation between apparent jet
speed and absolute optical magnitude does not seem to be as significant as the
one observed between apparent jet speed and extended radio luminosity (Table 2).
This could in principle undermine the suggestion that the optical luminosity is
a better indicator of jet kinetic power than extended radio luminosity.
However, since the optical luminosity is more likely to be affected by
strong variability, than the extended radio luminosity, the lack of a strong 
correlation can perhaps be accounted for, in these non-contemporaneous radio and optical
observations. Nevertheless, this is an important result to bear in mind, that
requires further testing.

\subsection{Environmental Effects}
\begin{figure}[ht]
\centerline{
\includegraphics[width=9cm]{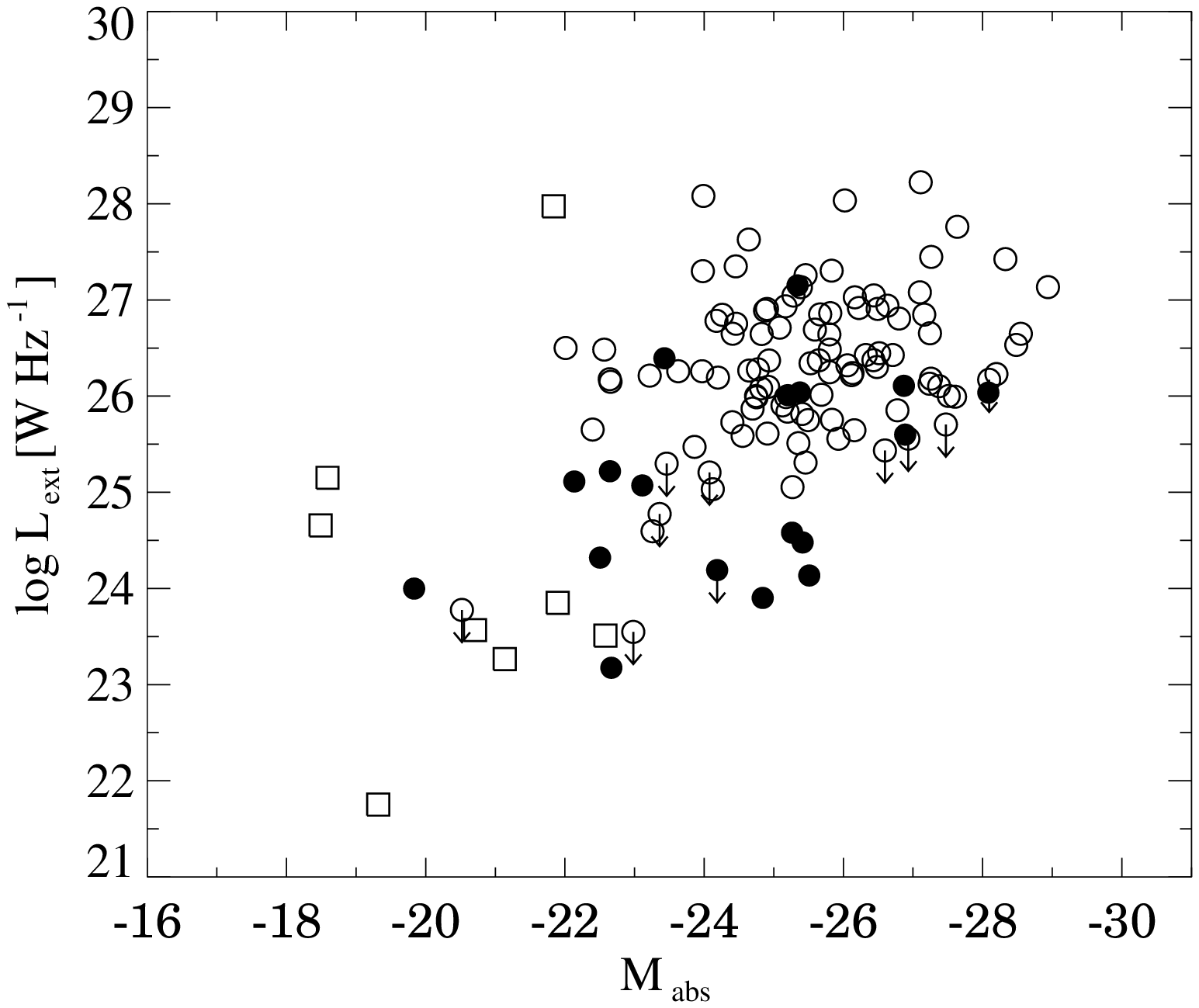}
\includegraphics[width=9cm]{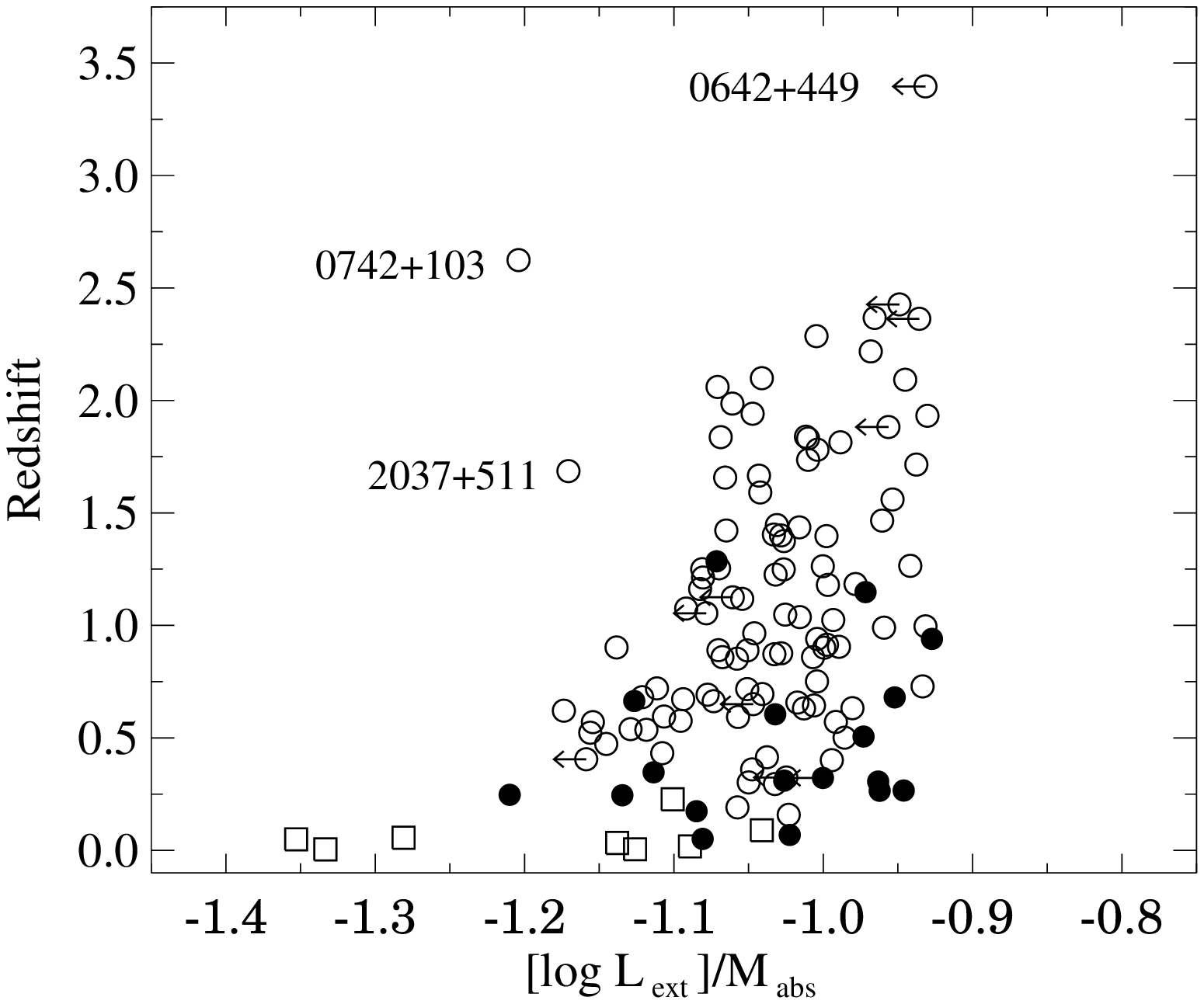}}
\caption{\small (Left) 1.4~GHz extended radio luminosity versus the absolute
optical magnitude. (Right) The environment indicator versus redshift. Open and filled 
circles denote quasars and BL~Lacs, respectively, while open squares denote 
radio galaxies. Core-only sources are represented as upper limits.}
\label{fig:opt}
\end{figure}

It has been suggested that apart from projection and relativistic beaming 
effects, the complex radio structures in quasars could arise due to 
interactions with the surrounding medium \citep{PearsonReadhead88,Barthel88}.
As is observed in Figure~\ref{fig:opt}, the extended radio luminosity
appears to be correlated with absolute optical luminosity. 
However, a partial regression analysis shows that the linear correlation 
becomes weak or non-existent, after the effects 
of luminosity distance are removed (for quasars, $r_{XY.Z}$ = $-$0.169, $t$ 
statistic = $-$1.62, $p$ = 0.1088; for BL Lacs, $r_{XY.Z}$ = 0.079, $t$ statistic
= 0.31, $p$ = 0.7608). The lack of a strong correlation may again be attributable 
to strong optical variability in these blazars.
If the extended radio luminosity is indeed affected by interaction with the 
kiloparsec-scale environment, and the optical luminosity is closely related to 
the AGN power, the ratio [log~$L_{ext}]/M_{abs}$, can serve as a probe for
environmental effects on kiloparsec-scales 
\citep[as suggested by][]{WillsBrotherton95}.
Henceforth we will refer to this ratio as the ``environment indicator'' (we will
drop the ``log'' term in the ratio for convenience, in the following text).

A higher value of the ratio could indicate greater jet-medium interaction,
(as say, from an asymetrically dense source environment).
Dense environments can decrease expansion losses in the source, but 
increase radiative losses, making the sources brighter at low radio frequencies 
\citep[e.g., in Cygnus A, see][]{Barthel96}.
Note that the optical magnitude in blazars is expected to be related to the
AGN (accretion disk or jet), rather than to starlight.
This might not be true for the nearby radio galaxies, where
starlight might indeed be a significant contributor to the optical magnitude.
The $L_{ext}/M_{abs}$ ratio might therefore not be a good environment proxy
for the radio galaxies. We have therefore excluded radio galaxies
from the correlation tests related to the environment proxy indicator. 

We plot the environment proxy ratio with respect to redshift 
in the right hand panel of Figure~\ref{fig:opt} and 
find a strong positive correlation. 
The ``alignment effect'' between the radio source axis and 
the emission line region in high redshift radio {\it sources} demonstrates
that local galactic asymmetries (on scales of several kiloparsecs)
increase with redshift \citep{McCarthy93,Best99}.
This result therefore, is consistent with the suggestion of the 
environmental dependence of the extended luminosity.
We note that the BL Lacs considered alone fail to show a strong
correlation (Table~\ref{tabcorrel}), which could be a consequence of an 
inhomogeneous parent population, as discussed in \S3.3.  

\begin{figure}[ht]
\centerline{
\includegraphics[width=9cm]{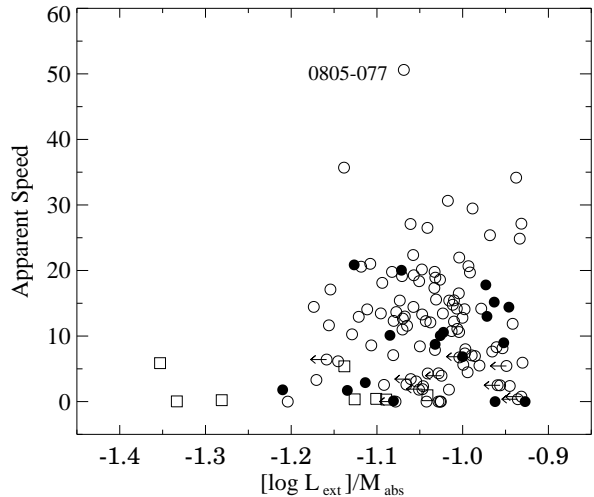}}
\caption{\small Apparent parsec-scale jet speed versus the environment indicator.
Open and filled 
circles denote quasars and BL~Lacs, respectively, while open squares denote 
radio galaxies.}
\label{fig:envbeta}
\end{figure}
\begin{figure}[ht]
\centerline{
\includegraphics[width=9cm]{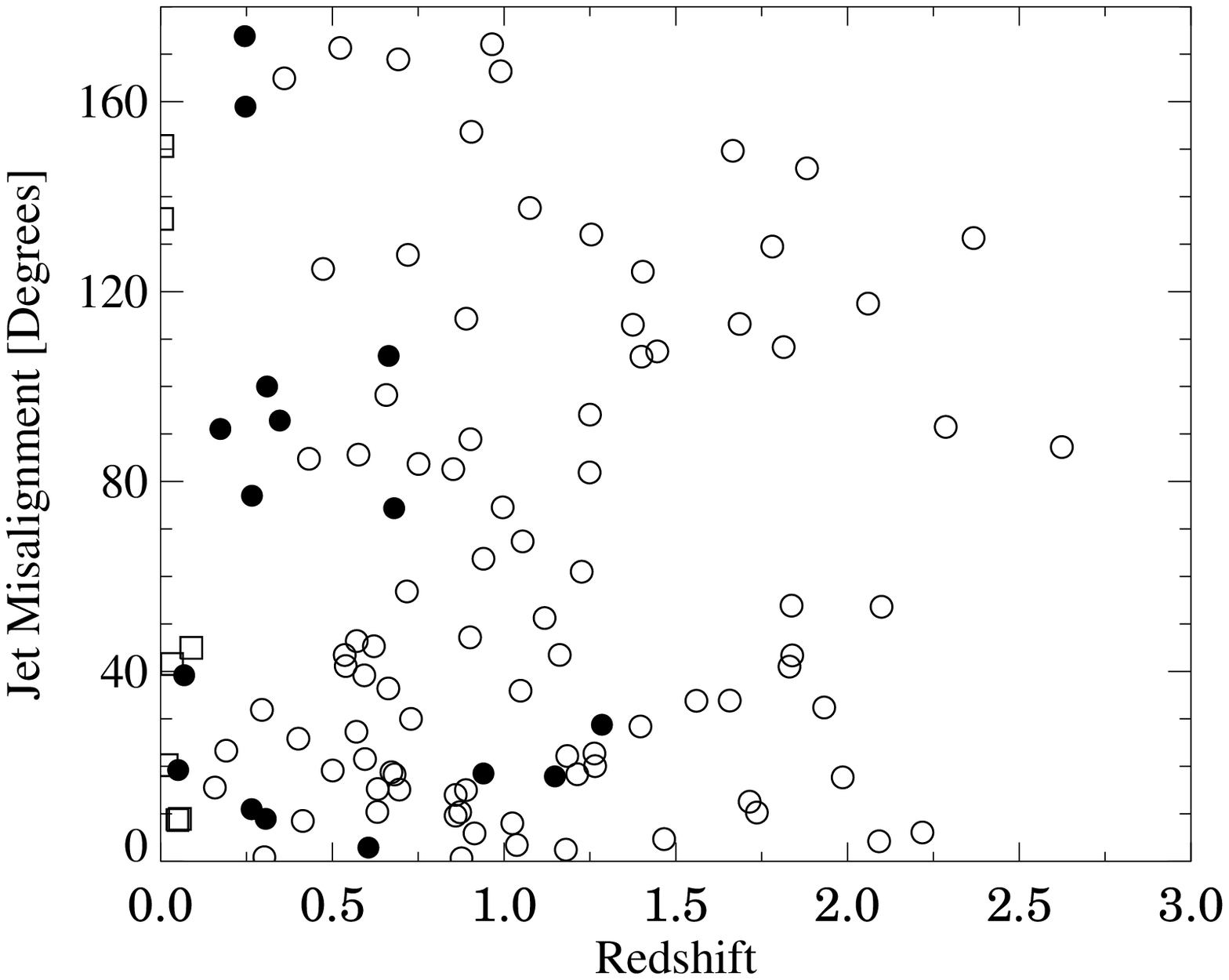}
\includegraphics[width=9cm]{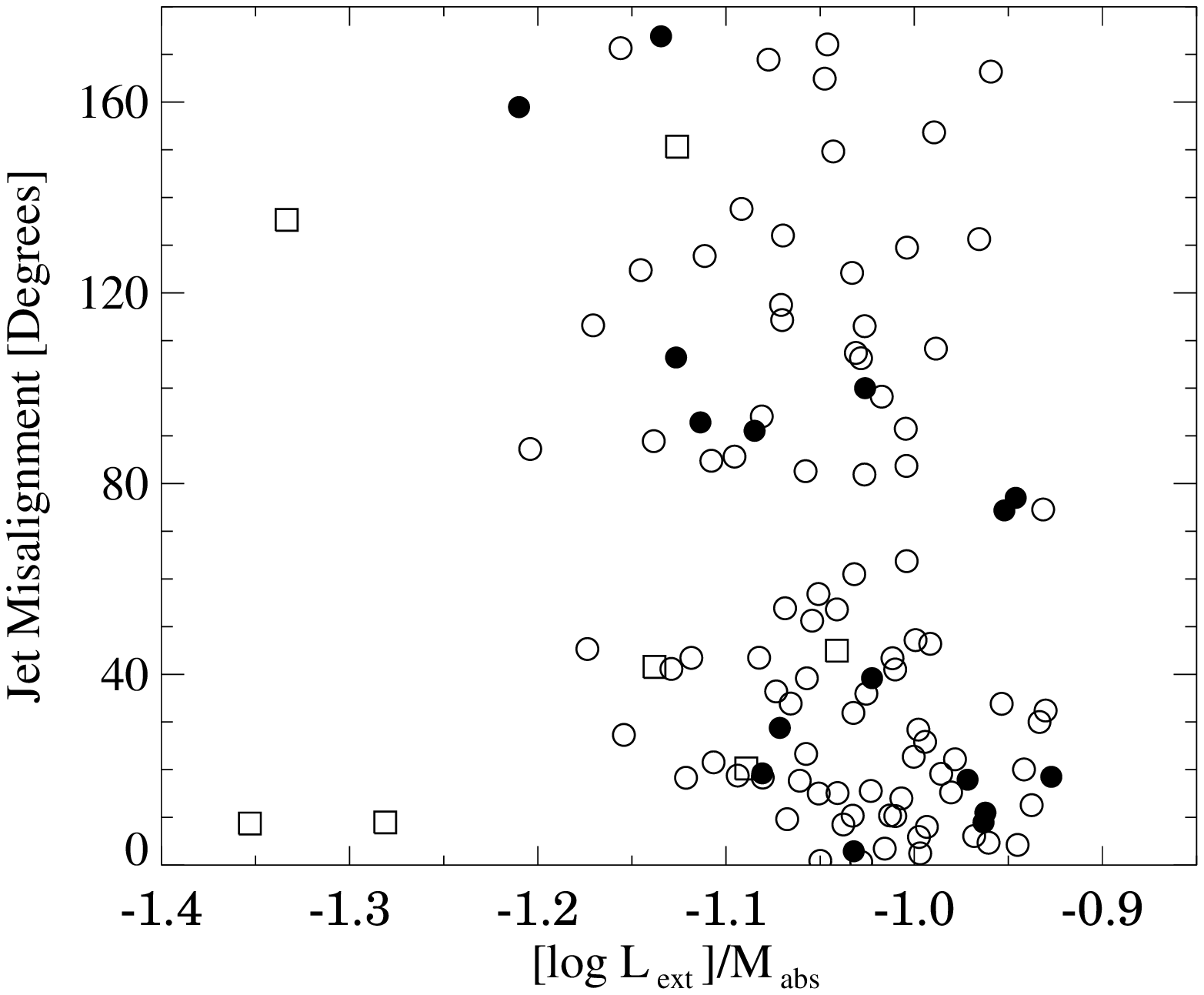}}
\caption{\small (Left) Jet misalignment versus the environment indicator.
(Right) Misalignment angle versus redshift.
Open and filled 
circles denote quasars and BL~Lacs, respectively, while open squares denote 
radio galaxies.}
\label{fig:DeltaBeta}
\end{figure}

We find that while the parsec-scale jet speeds appear to be tightly 
correlated with the extended radio luminosity (Fig.~\ref{fig:beta}), 
and weakly correlated with the optical luminosity, they seem
to not show any correlation with the kpc-scale environment indicator
(Fig.~\ref{fig:envbeta}). This could imply that the
parsec-scale jet speeds are not driven by the kiloparsec-scale environment, but 
rather by factors intrinsic to the AGN, or the AGN's local parsec-scale 
environment. For instance, the jet speeds could be related to the
black hole spins \citep[e.g.,][]{Meier99}. 
However, like other results pertaining to optical luminosity, this
lack of a correlation needs further testing with contemporaneous optical and
radio data.

We find that the parsec-to-kiloparsec jet misalignment is not correlated
with redshift (Fig.~\ref{fig:DeltaBeta}). This goes against the idea 
that the bending of the jets is influenced by interaction with the medium; 
an effect which should get more prominent with increasing redshift. This result 
is consistent with the suggestions of \citet{Hintzen83} and \citet{Appl96}. 
We note that \citet{Hintzen83} used the
kiloparsec-scale jet-to-counterjet misalignment in their study. \citet{Kharb08a}
found a similar lack of correlation between jet-to-counterjet misalignment
and redshift for an FRII radio galaxy sample. 
Thus, the jet misalignment could also be related 
to factors closely associated with the AGN itself. For instance, the presence 
of binary black holes or kicks imparted to the black hole via a black hole merger, 
could switch the ejection direction on a short timescale 
\citep[e.g.,][]{Begelman80,Merritt02}. Black hole re-alignment due to a 
warped accretion disk could also give rise to jet precession \citep{Pringle97}. 
We note that a few sources have morphologies resembling fast, precessing 
jets (``banana'' shaped) as modelled by \citet{Gower82}. 
Alternatively, the jet misalignment could be influenced by the AGN's local
parsec-scale environment.

The jet misalignment is inversely correlated with the kpc-scale environment 
indicator (see right hand panel of Fig.~\ref{fig:DeltaBeta}). The inverse relation
suggests that jets with smaller misalignments
($i.e.,$ straighter jets) show signatures of greater environmental effects
on their lobe emission.
This could either be suggestive of a uniformly dense confining medium on
different spatial scales around the source, or, of projection effects in
the sample, which could boost the optical emission and decrease the
$L_{ext}/M_{abs}$ ratio in sources with larger jet misalignments. 
If the latter is true, then this is an important caveat $-$ the environment
indicator could also be affected by source orientation in blazars. However, 
the orientation-dependence alone cannot satisfactorily explain the
correlation between the environment indicator and redshift (Fig.~\ref{fig:opt}).

Many MOJAVE blazars exhibit radio structures resembling those of {\it wide angle
tail} radio galaxies. Observations have indicated that WAT quasars could be members 
of rich galaxy clusters \citep[e.g.,][]{Hintzen83,Harris83,Blanton00}.
Using the NASA/IPAC extragalatic database (NED) we found that nearly $7\%$ of the 
MOJAVE sources have a galaxy cluster $\sim5\arcmin$ away, while 
$14\%$ have a galaxy cluster $\sim10\arcmin-15\arcmin$ away. 
While this fraction appears to be small, we must keep in mind that most of the 
blazars are at high redshifts, while the cluster information is highly 
incomplete even at redshifts $z>0.1$ \citep[e.g.,][]{Ebeling98,Bohringer04}.
Thus the possibility remains that a large number of MOJAVE blazars inhabit 
galaxy clusters. This suggestion has interesting ramifications, as 
discussed below. Furthermore, this could help us understand some of the
results discussed in this section.

\subsection{Relation Between Radio Power and Emission-line Luminosity}
\citet{RawlingsSaunders91} demonstrated a tight correlation between emission-line 
luminosity and total jet kinetic power in 
a sample of radio-loud AGNs. This correlation was tighter than the one
observed between emission-line luminosity and radio luminosity. \citet{Landt06} 
discovered a large number of ``blue quasars'' with strong emission lines,
but relatively low radio powers in the Deep X-ray Radio Blazar Survey (DXRBS) and
the RGB samples, and suggested that this went against the close relationship
between emission-line and jet power. These and other observations 
\citep[e.g.,][]{Xu99} could be suggesting that the mechanism for the production of
the optical-UV photons that ionise the emission-line clouds is decoupled
from the mechanism that produces powerful radio jets.
Many MOJAVE blazars also seem to not follow the relation between 
emission-line and radio luminosity; that is, some quasars and BL~Lacs 
have relatively low and high radio powers, respectively, and the suggestion 
of decoupled mechanisms might hold true. However, an alternate
hypothesis by which their emission-line luminosity could still be correlated with
their intrinsic jet kinetic power, could come about if the outlying BL~Lac 
objects
were present in dense confining environments, while the outlying 
quasars were old,
field sources. Confinement would increase the synchrotron radiative losses due to the 
amplification of the magnetic field strength, making the sources brighter
\citep[see][]{Barthel96}, while expansion losses and gradual radiative
losses could
presumably reduce the low frequency radio emission in the old, field quasars.

It has been suggested that a large fraction ($\sim$50\%) of the high redshift
radio sources could be GPS sources \citep[e.g.,][]{ODea91}, which are
likely to be present in confined environments. One of the highest 
redshift MOJAVE sources, 0742+103 ($z=2.624$), is a GPS quasar.
There are several other MOJAVE sources that have peaked spectra, but
do not show the low-variability or the double-lobed parsec-scale structures that are
characteristic of the GPS class. 
These could be ``masquerading GPS" sources, whose overall spectrum happens to 
be dominated by an unusually bright feature in the jet, located downstream 
from the core.

\subsection{Blazar Division Based on Radio Power}
In order to gain more insight into the primary factors that drive the FR
dichotomy, in this section we disregard the emission-line division 
(i.e., quasars, BL~Lacs), 
and divide sources into FRI, FRI/II and FRII classes based solely on 
their 1.4~GHz extended powers (following Fig.~\ref{fig:z}).
Figure~\ref{fig:Env} plots the environment indicator, $L_{ext}/M_{abs}$, 
with respect to redshift, using the new classification. A quick comparison with 
Fig.~\ref{fig:opt} reveals 
that many of the low redshift quasars fall into the intermediate 
FRI/II class. The low redshifts of these sources discounts the possibility
of a $(1+z)^4$ surface brightness dimming effect in them, 
which could potentially reduce their extended luminosity, making them fall
in the intermediate luminosity class. This finding is consistent with the 
discovery of ``blue quasars'' by \citet{Landt06},
which have radio powers and spectral energy distributions (SEDs) 
similar to the high energy peaked BL Lacs.

\begin{figure}[ht]
\centerline{
\includegraphics[width=9cm]{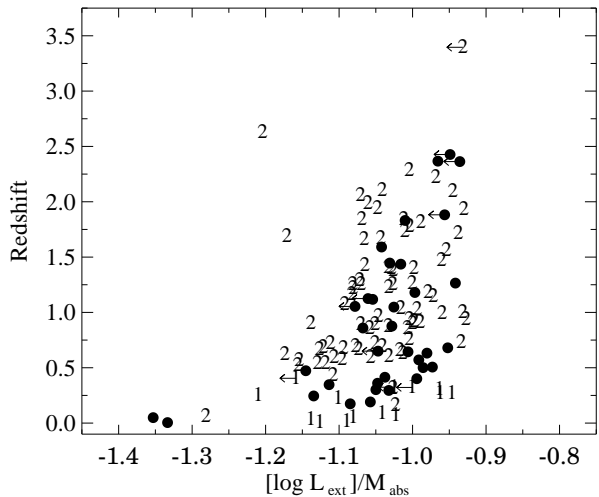}}
\caption{\small The environment indicator versus redshift, with sources 
divided on the basis of extended radio luminosity into FRI (=1), FRII (=2), 
and FRI/II (=filled circles).} 
\label{fig:Env}
\end{figure}

Most beamed FRIs lie at redshifts below 0.5 in Fig.~\ref{fig:Env}. 
This could be a limitation of the flux limited sample.
Interestingly, the FRI/II sources lie at all redshift ranges,
and throughout the $L_{ext}/M_{abs}$ distribution. This suggests that
these ``intermediate'' luminosity sources are not produced as a result of different
environmental conditions compared to, say, other FRIIs. 
Similarly, sources divided on the basis of the presence or absence of 
hot spots, show no discernable difference in the $L_{ext}/M_{abs}$ vs. $z$ plane.
It is important to note that environmental radio boosting, as
mentioned in \S4.4 \citep[also][]{Barthel96} could be instrumental in
blurring the Fanaroff-Riley dividing line in the MOJAVE blazars.

Lastly, the orientation indicator, $R_v$, seems to be 
correlated with misalignment only for FRIIs, but not for FRIs or FRI/IIs
(Table~\ref{tabcorrel}).
This implies that projection is playing a significant role in the jet 
misalignment of FRII sources, but the jets are intrinsically more 
distorted in FRIs and FRI/IIs. A corollary to this is that the differences between 
FRIs and FRIIs could be related to factors affecting the black hole spin direction, 
since we have previously concluded that jet mislignments appear to be
largely unrelated to kiloparsec-scale environmental effects. 

\subsection{Extremely Misaligned Sources}
Six MOJAVE sources ($viz.,$ 0224+671, 0814+425, 1219+044, 1510$-$089, 
1828+487, 2145+067) have jet misalignments greater than 160$\degr$. 
Such extremely misaligned  sources could be bent jets viewed nearly head-on, as 
suggested for the gamma-ray blazar PKS 1510$-$089 \citep{Homan02}. 
An alternate scenario could be that some of these misaligned sources are ``hybrid'' 
morphology sources \citep[e.g.,][]{Gopal-Krishna00}. 
Three of six sources with large misalignments ($viz.,$ 0224+671, 1219+044, 
1510$-$089) indeed seem to possess radio morphologies that could be classified 
as ``hybrid''.
The large misalignment then makes sense: the VLBI jet (used to determine the 
parsec-scale jet position angle) could be on the FRI side, while the
sole hot spot (used to determine the kiloparsec-scale jet position angle) 
could be on the FRII side. 

One interesting source that shows extreme misalignment ($\sim135\degr$) is M87
(1228+126). M87 appears to exhibit a weak hot spot on the counterjet
side \citep[e.g.,][]{Owen90,Sparks92}. Since the brightest emission on the 
counterjet side is at the extremity of the radio lobe (like in FRIIs), M87 could be 
classified as a hybrid radio morphology source. 
Its radio luminosity ($L_{178}=1\times10^{25}$ W~m$^{-2}$) also places it close
to the FRI/FRII division \citep{Biretta99}.
\citet{Kovalev07} have 
observed a very short, weak radio counterjet on parsec-scales in M87, but 
perhaps that is just a slower outer sheath that quickly terminates.

If indeed some of the extremely misaligned sources are ``hybrid'' sources, the
lack of a correlation between jet misalignment and environment appears to contradict 
the suggestion that asymmetries in galactic environments, which increase with 
redshift \citep[e.g.,][]{Best99}, give rise to hybrid morphology sources
\citep[e.g.,][]{Gopal-Krishna00,Miller09}.
Moreover, the fraction of such sources would be higher in the MOJAVE ($6\%-8\%$), 
than that reported for the FIRST survey \citep[1\%,][]{Gawronski06}, 
suggesting that hybrid sources may not be as rare as previously supposed.
It is also interesting to note that not many MOJAVE ``hybrid'' morphology
sources fall in the FRI/II luminosity range, but rather have FRII luminosities.
In other words, the MOJAVE ``hybrid'' morphology sources are not necessarily
``hybrid'' in terms of radio power, but mostly fall into the FRII luminosity class.
Chandra X-ray observations of these sources could resolve some of these 
apparent contradictions. 

\subsection{Core-only Sources}
About 7\% of the MOJAVE sources do not show any extended radio emission in 
our 1.4 GHz VLA A-array images.
However the ratio of their 5~GHz radio flux density to their B-band 
optical flux density is of the order of a few hundred, placing them firmly in 
the radio-loud AGN class \citep[e.g.,][]{Kellermann89}.
The dynamic range achieved in the radio images of these sources varied from 
about 6000:1
($e.g.,$ 0955+476) to 30,000:1 ($e.g.,$ 0202+149), being typically of the
order of 10,000:1. While some sources, like 1749+096, failed to reveal any
extended emission in a number of different datasets 
\citep[see also][]{Rector01}, we did detect faint halo-like emission
around 1548+056, which was listed as a core-only source by \citet{Murphy93}.    
Note that observations with the VLA A-array-only are likely to miss
very diffuse radio structure, due to the lack of short baselines. 
We believe therefore that these core-only sources probably have associated
faint emission which require more sensitive observations to detect.

The observed fraction of core-only sources in MOJAVE is smaller than 
that reported for other quasar surveys \citep[$20\%$,][]{Gower84,Padrielli88}.
While the core-only sources could have been affected by the redshift surface 
brightness dimming effect, our discussion in the previous section argues
against this conjecture. 
Alternatively, these sources could just be normal quasars at very small angles to 
line of sight, so that the VLA resolution is insufficient to delineate 
the various components. This appears to be the case in 1038+064,
and 1417+385, where a two Gaussian component model fits the ``core''
emission better than a single one.
The hypothesis that at least some of these quasars 
could be beamed FRIs also cannot be ruled out \citep[e.g.,][]{Heywood07}. 

\subsection{Optical Polarization Subclasses}
Low optical polarization radio quasars (LPRQs) consistently reveal
core optical fractional polarization, $m_{opt}<3\%$, while the high optical
polarization quasars (HPQs) routinely reveal highly polarized
optical cores with $m_{opt}\ge3\%$ \citep{AngelStockman80}. The classification
of the MOJAVE sources into these categories is listed in Table~\ref{tabsample},
and were obtained from \citet{Impey91} and \citet{Lister00}.  
In all, there are 38 HPQs and 16 LPRQs in the MOJAVE sample.
We find no statistically significant differences in the core or
extended radio luminosity between LPRQs and HPQs. In \citet{Kharb08}
we had noted a difference in the integrated (single-dish) 1.4~GHz radio 
luminosity between LPRQs and HPQs for redshifts less than one. We do not
find a difference in the total radio luminosity using the present higher resolution 
VLA observations. Furthermore, we do not find a statistically significant 
difference in jet misalignment between LPRQs and HPQs.
The parsec-scale apparent jet speeds of HPQs, however, are systematically
higher than those in LPRQs (Kolmogorov-Smirnov test probability that
the speeds in HPQs and LPRQs are similar, is $p$=0.03).
Additional optical polarimetry information on
the sample is needed to further investigate these trends.

\section{SUMMARY AND CONCLUSIONS}
\begin{enumerate}
\item We describe the results from a study of the extended emission
at 1.4~GHz in the MOJAVE sample of 135 blazars. New VLA A-array images of 
six MOJAVE quasars, and previously unpublished VLA A-array archival images of 
21 blazars, are presented. The kiloparsec-scale structures
are varied, with many sources displaying peculiar, distorted morphologies. 
Many blazar jets show parsec-to-kiloparsec-scale jet misalignments greater 
than $90\degr$. These characteristics have been reported in other quasar 
surveys as well.
The $90\degr$ bump in the jet misalignment distribution
(of which we observe a weak signature) has been suggested to be a result of  
low-pitch helical parsec-scale jets in the literature. 

\item A substantial number of MOJAVE quasars ($\sim22\%$) and 
BL~Lacs ($\sim23\%$) display radio powers that are intermediate between FRIs 
and FRIIs. Many BL~Lacs have extended luminosities ($\sim27\%$) and hot spots 
($\sim60\%$) like quasars. The hypothesis that at least some of the 
core-only quasars are in fact FRI quasars, cannot be discarded. 
In terms of radio properties alone, it is difficult to draw a sharp dividing
line between BL~Lacs and quasars, in the MOJAVE sample.
This could be a result of the MOJAVE selection effects which naturally pick
sources with a large range in intrinsic radio power (low power quasars and high
power BL Lacs), and preferentially bent jets.
While the quasars and BL Lacs display a smooth continuation in all their properties,
the correlation test results differ sometimes when they are considered separately.
This can be understood if the BL Lacs did not constitute a homogeneous population
like the quasars, but had both FRI and FRII radio galaxies as their parent population.
These findings challenge the simple radio-loud unified scheme, which links FRI 
sources to BL~Lac objects and FRIIs to quasars.

\item We find that there is a significant correlation between extended radio
luminosity and apparent parsec-scale jet speeds. This implies that most 
quasars have faster jets than most BL~Lac objects. The large overlap between many 
properties of quasars and BL~Lacs (e.g., 1.4~GHz radio power, parsec-scale jet 
speeds) however suggests that, at least in terms of radio jet properties, 
the distinction between these two AGN classes, at an
emission-line equivalent width of $5\AA$, is essentially an arbitrary one.

\item These observations suggest that the mechanism for the production 
of optical-UV photons that ionise the emission-line clouds is decoupled from the 
mechanism that produces powerful radio jets \citep[e.g.,][]{Xu99,Landt06}. An
alternate hypothesis could be that, 
BL~Lacs with high radio powers are present in dense confining 
environments, which would increase the radiative losses in them and make them
brighter, while the radio quasars with low radio powers are old, field sources.
Environmental radio boosting \citep{Barthel96} could also explain why the
Fanaroff-Riley dividing line is blurred in the MOJAVE sample.
X-ray cluster studies could be used to test these ideas.

\item The ratio of the radio core luminosity to the $k$-corrected optical
luminosity ($R_v$) appears to be a better indicator of jet orientation than the
traditionally used radio core prominence parameter ($R_c$). This seems to be a 
consequence of the environmental contribution to the extended radio luminosity,
used in $R_c$. Trends with $R_v$ reveal that even though the sample comprises
largely of blazars, there is a reasonable range of orientations present in 
the sample.

\item Trends with the ``environment'' proxy indicator, $L_{ext}/M_{abs}$, 
reveal that jet speeds seem to not depend on the environment on 
kiloparsec scales, while the jet misalignments seem to be inversely correlated. 
The jet misalignments are also not correlated with redshift.
It appears that the parsec-to-kiloparsec jet misalignment, the parsec-scale 
jet speeds and to an extent the extended emission (which is related to jet speed) 
are controlled by factors intrinsic to the AGN. Black hole spins could dictate
the jet speeds, while the presence of binary black holes, or kicks imparted to 
black holes via black hole mergers could influence the jet direction. This is
consistent with radio morphologies similar to precessing jet models of
\citet{Gower82} present in the MOJAVE sample, and the signature of the 90$\degr$ 
bump in the jet misalignment distribution, which is attributed to low-pitch helical
parsec-scale jets.

\item If some of the highly misaligned sources (jet misalignment $\sim180\degr$) 
are ``hybrid'' FRI+II morphology sources, then 
the fraction of such sources is higher in the MOJAVE survey ($6\%-8\%$), than that
reported for the FIRST survey \citep[1\%,][]{Gawronski06}. Furthermore, the
lack of a correlation between jet misalignment and the environmental 
indicator would appear to contradict the suggestion that hybrid morphology is a 
result of the jet interacting with an asymmetric environment.
X-ray observations of hybrid sources are needed to resolve these issues.

\item About $7\%$ of the MOJAVE quasars show no extended radio emission. 
We expect more sensitive radio observations to detect faint emission in these 
sources, as we have detected in the case of 1548+056, previously classified as
a core-only quasar. The hypothesis that at least some of these quasars 
could be beamed FRIs cannot be ruled out. 

\item While the extended radio power or jet misalignments do not show
statistically significant differences between the two quasar optical polarization 
subclasses, $viz.,$ LPRQs
and HPQs, the parsec-scale apparent jet speeds of HPQs are systematically
higher than those in LPRQs. 
\end{enumerate}

\begin{deluxetable}{llllccllllllllllllllllll}
\tabletypesize{\tiny}
\tablecaption{The MOJAVE sample}
\tablewidth{0pt}
\tablehead{
\colhead{Source}&\colhead{z}&\colhead{$V$}&\colhead{Type}&\colhead{Opt}&\colhead{$I_{peak}$}&\colhead{$I_{rms}$}&\colhead{Freq}&\colhead{$\beta_{app}$}&\colhead{$S_{core}$}&\colhead{$S_{ext}$}&\colhead{Radio}&\colhead{PcjetPA}&\colhead{KpcjetPA}&\colhead{Ref}\\
\colhead{name}&\colhead{}&\colhead{mag}&\colhead{}&\colhead{Pol}&\colhead{Jy~beam$^{-1}$}&\colhead{Jy~beam$^{-1}$}&\colhead{GHz}&\colhead{}&\colhead{Jy}&\colhead{mJy}&\colhead{Morph}&\colhead{Degree}&\colhead{Degree}&\colhead{}\\
\colhead{(1)}&\colhead{(2)}&\colhead{(3)}&\colhead{(4)}&\colhead{(5)}&\colhead{(6)}&\colhead{(7)}&\colhead{(8)}&\colhead{(9)}&\colhead{(10)}&\colhead{(11)}&\colhead{(12)}&\colhead{(13)}&\colhead{(14)}&\colhead{(15)}}
\startdata
0003$-$066    &   0.347  &  18.50 &  B  &  Y   &2.605 &1.6$\times10^{-4}$& 1.4000& 2.89     & 2.66 & 43.9     & 2              & 282 &15   &    AL634$\ast$   \\
0007$+$106    &  0.0893  &  15.40 &  G  & ...  &0.075 &4.0$\times10^{-5}$& 1.4000& 0.97     & 0.08 & 17.6     & 2hs            & 284 &239  &    AL634$\ast$   \\
0016$+$731    &   1.781  &  19.00 &  Q  &  N   &0.398 &4.2$\times10^{-4}$& 1.4000& 6.74     & 0.40 & 7.8      & 1hs            & 132 &2    &    AL634$\ast$   \\
0048$-$097    &    ....  &  17.44 &  B  &  Y   &0.562 &1.8$\times10^{-4}$& 1.4250& ....     & 0.57 & 139.7    & 1hs+2, 2hs     & 337 &190  &    AB1141        \\
0059$+$581    &   0.644  &  17.30 &  Q  & ...  &1.554 &1.1$\times10^{-4}$& 1.4000& 11.08    & 1.57 & 19.4     & ch+1           & 236 &.... &    AL634$\ast$   \\
0106$+$013    &   2.099  &  18.39 &  Q  &  Y   &2.760 &1.4$\times10^{-4}$& 1.4000& 26.50    & 2.81 & 530.6    & 1hs, 2          & 238 &184  &    AL634$\ast$   \\
0109$+$224    &   0.265  &  15.66 &  B  &  Y   &0.361 &7.8$\times10^{-5}$& 1.4000& ....     & 0.36 & 3.9      & 1	       & 82  &93   &    AL634$\ast$   \\
0119$+$115    &    0.57  &  19.70 &  Q  & ...  &1.188 &9.8$\times10^{-5}$& 1.4000& 17.09    & 1.24 & 113.7    & 1hs, ch         & 6   &33   &    AL634$\ast$   \\
0133$+$476    &   0.859  &  19.50 &  Q  &  Y   &1.872 &9.6$\times10^{-5}$& 1.4000& 12.98    & 1.88 & 8.7      & 1hs             & 331 &340  &    AL634$\ast$   \\
0202$+$149$^c$&   0.405  &  21.00 &  Q  &  Y   &3.808 &1.4$\times10^{-4}$& 1.4000& 6.41	    & 3.86 & 1.1      & c	       & 319 &.... &    AL634$\ast$   \\
0202$+$319    &   1.466  &  17.40 &  Q  &  N   &0.648 &8.4$\times10^{-5}$& 1.4000& 8.30     & 0.65 & 11.7     & 1hs             & 357 &2    &    AL634$\ast$   \\
0212$+$735    &   2.367  &  19.00 &  Q  &  Y   &2.458 &1.0$\times10^{-4}$& 1.4000& 7.63     & 2.47 & 1.7      & 1hs, 2hs         & 122 &253  &    AL634$\ast$   \\
0215$+$015    &   1.715  &  16.09 &  B  &  Y   &0.416 &6.9$\times10^{-5}$& 1.4000& 34.16    & 0.45 & 71.1     & 2hs             & 104 &92   &    AL634$\ast$   \\
0224$+$671    &   0.523  &  19.50 &  Q  & ...  &1.476 &1.0$\times10^{-4}$& 1.4000& 11.63    & 1.48 & 149.2    & 2hs             & 355 &184  &    AL634$\ast$   \\
0234$+$285    &   1.213  &  19.30 &  Q  &  Y   &2.274 &1.0$\times10^{-4}$& 1.4000& 12.31    & 2.33 & 99.9     & 1hs             & 349 &331  &    AL634$\ast$   \\
0235$+$164    &    0.94  &  15.50 &  B  &  Y   &1.501 &5.8$\times10^{-5}$& 1.4000& ....     & 1.51 & 25.5     & 1hs+2           & 325 &343  &    AL634$\ast$   \\
0238$-$084    &   0.005  &  12.31 &  G  & ...  &1.091 &6.0$\times10^{-5}$& 1.4000& 0.34     & 1.11 & 111.3    & 2              & 66  &275  &    AL634$\ast$   \\
0300$+$470    &    ....  &  16.95 &  B  &  Y   &1.166 &5.7$\times10^{-5}$& 1.4000& ....     & 1.18 & 60.1     & 1, ch          & 148 &225  &    AL634$\ast$   \\
0316$+$413    &  0.0176  &  12.48 &  G  &  N   &20.76 &1.0$\times10^{-3}$& 1.3121& 0.31     & 21.40& 1103.0   & 2hs             & 178 &158  &    BT024         \\
0333$+$321    &   1.263  &  17.50 &  Q  &  N   &2.994 &1.0$\times10^{-4}$& 1.4000& 12.78    & 3.02 & 71.8     & 1hs             & 126 &149  &    AL634$\ast$   \\
0336$-$019    &   0.852  &  18.41 &  Q  &  Y   &2.908 &1.5$\times10^{-4}$& 1.4000& 22.36    & 2.92 & 70.3     & 1hs             & 67  &344  &    AL634$\ast$   \\
0403$-$132    &   0.571  &  17.09 &  Q  &  Y   &4.043 &3.3$\times10^{-4}$& 1.4000& 19.69    & 4.33 & 9.1      & 1hs+2           & 155 &202  &    AL634$\ast$   \\
0415$+$379    &  0.0491  &  18.05 &  G  & ...  &0.542 &5.7$\times10^{-4}$& 1.4899& 5.86     & 0.57 & 2700.0   & 2hs	       & 71  &62   &    AR102         \\
0420$-$014    &   0.914  &  17.00 &  Q  &  Y   &2.887 &1.0$\times10^{-4}$& 1.4000& 7.34     & 2.91 & 70.2     & 2hs, 1hs+2       & 183 &189  &    AL634$\ast$   \\
0422$+$004    &    ....  &  16.98 &  B  &  Y   &1.083 &1.3$\times10^{-4}$& 1.4000& ....	    & 1.09 & 6.1      & 2hs             & 357 &200  &    AL634$\ast$   \\
0430$+$052    &   0.033  &  15.05 &  G  & ...  &2.845 &1.7$\times10^{-4}$& 1.4649& 5.38     & 2.95 & 159.5    & 1, 2           & 252 &293  &    AB379         \\
0446$+$112    &    ....  &  20.00 &  U  & ...  &1.527 &1.2$\times10^{-4}$& 1.4000& ....     & 1.56 & 15.4     & 1hs             & 103 &207  &    AL634$\ast$   \\
0458$-$020    &   2.286  &  18.06 &  Q  &  Y   &1.629 &3.8$\times10^{-4}$& 1.4899& 16.51    & 1.66 & 148.1    & 1hs, 2hs         & 321 &230  &    AN047         \\
0528$+$134    &    2.06  &  20.00 &  Q  & ...  &2.204 &1.2$\times10^{-4}$& 1.4000& 19.14    & 2.24 & 60.1     & 2hs, 1hs+2       & 19  &261  &    AL634$\ast$   \\
0529$+$075    &   1.254  &  19.00 &  Q  & ...  &1.518 &5.7$\times10^{-5}$& 1.4000& 12.66    & 1.54 & 126.8    & 2hs, 1hs+2       & 353 &221  &    AL634$\ast$   \\
0529$+$483    &   1.162  &  19.90 &  Q  & ...  &0.641 &6.7$\times10^{-5}$& 1.4000& 19.78    & 0.65 & 21.5     & 1hs+2           & 32  &75   &    AL634$\ast$   \\
0552$+$398$^c$&   2.363  &  18.30 &  Q  & ...  &1.539 &1.0$\times10^{-4}$& 1.4000& 0.36     & 1.55 & 1.2      & c	       & 290 &.... &    AL634$\ast$   \\
0605$-$085    &   0.872  &  17.60 &  Q  & ...  &2.344 &2.8$\times10^{-4}$& 1.4250& 19.79    & 1.20 & 123.9    & 1hs             & 109 &99   &    AD298         \\
0607$-$157$^c$&   0.324  &  18.00 &  Q  & ...  &3.011 &1.4$\times10^{-4}$& 1.4000& 3.94     & 3.02 & 1.1      & c	       & 52  &.... &    AL634$\ast$   \\
0642$+$449$^c$&   3.396  &  18.49 &  Q  & ...  &0.644 &7.5$\times10^{-5}$& 1.4000& 0.75	    & 0.65 & 1.4      & c	       & 90  &.... &    AL634$\ast$   \\
0648$-$165    &    ....  &   .... &  U  & ...  &2.085 &1.4$\times10^{-4}$& 1.4000& ....     & 2.12 & 11.2     & c, 1           & 273 &272  &    AL634$\ast$   \\
0716$+$714    &    0.31  &  15.50 &  B  &  Y   &0.645 &6.8$\times10^{-5}$& 1.4000& 10.06    & 0.69 & 376.4    & 2hs, 1hs+2       & 40  &300  &    AQ006         \\
0727$-$115    &   1.591  &  20.30 &  Q  & ...  &3.182 &9.4$\times10^{-5}$& 1.4250& ....     & 3.22 & 2.4      & c, 1hs          & 292 &.... &    AC874$\dagger$\\
0730$+$504    &    0.72  &  19.30 &  Q  & ...  &0.676 &6.0$\times10^{-5}$& 1.4250& 14.06    & 0.69 & 82.5     & 2hs	       & 210 &82   &    AC874$\dagger$\\
0735$+$178    &    ....  &  16.22 &  B  &  Y   &1.92  &0.7$\times10^{-3}$& 1.4650& ....	    & 1.91 & 20.6     & 1              & 77  &165  &    Murphy        \\
0736$+$017    &   0.191  &  16.47 &  Q  &  Y   &2.327 &3.2$\times10^{-4}$& 1.4899& 14.44    & 2.34 & 40.9     & 2hs	       & 273 &297  &    AA025         \\
0738$+$313    &   0.631  &  16.92 &  Q  & ...  &2.16  &0.4$\times10^{-3}$& 1.4650& 10.75    & 2.16 & 65.0     & 2hs	       & 156 &166  &    Murphy        \\
0742$+$103    &   2.624  &  24.00 &  Q  & ...  &3.600 &2.3$\times10^{-4}$& 1.4899& ....     & 3.62 & 5.8      & 1, 2           & 354 &81   &    AY012         \\
0748$+$126    &   0.889  &  18.70 &  Q  & ...  &1.43  &0.5$\times10^{-3}$& 1.4650& 18.36    & 1.43 & 27.0     & 2hs	       & 110 &125  &    Murphy        \\
0754$+$100    &   0.266  &  15.00 &  B  &  Y   &2.07  &0.4$\times10^{-3}$& 1.4650& 14.40    & 2.07 & 6.7      & 2hs	       & 10  &293  &    Murphy        \\
0804$+$499    &   1.436  &  19.20 &  Q  &  Y   &0.64  &0.3$\times10^{-3}$& 1.4650& 1.82     & 0.64 & 5.3      & 1hs	       & 134 &.... &    Murphy        \\
0805$-$077    &   1.837  &  19.80 &  Q  & ...  &1.439 &5.8$\times10^{-5}$& 1.4250& 50.60    & 1.58 & 59.8     & 1hs	       & 332 &278  &    AC874$\dagger$\\
0808$+$019    &   1.148  &  17.20 &  B  &  Y   &0.449 &9.1$\times10^{-5}$& 1.4899& 12.99    & 0.46 & 18.2     & 2hs	       & 182 &200  &    AA025         \\
0814$+$425    &   0.245  &  18.18 &  B  &  Y   &1.015 &3.1$\times10^{-4}$& 1.5159& 1.70     & 1.03 & 76.8     & 2hs	       & 136 &322  &    AK460         \\
0823$+$033    &   0.506  &  16.80 &  B  &  Y   &1.33  &0.4$\times10^{-3}$& 1.4650& 17.80    & 1.32 & 4.1      & 1hs	       & 15  &.... &    Murphy        \\
0827$+$243    &    0.94  &  17.26 &  Q  & ...  &0.743 &1.4$\times10^{-4}$& 1.5100& 21.97    & 0.76 & 62.7     & 1hs+2           & 135 &199  &    AV150         \\
0829$+$046    &   0.174  &  16.40 &  B  &  Y   &0.788 &1.9$\times10^{-4}$& 1.4000& 10.10    & 0.80 & 150.8    & 2, 1hs+2        & 60  &151  &    AG618         \\
0836$+$710    &   2.218  &  17.30 &  Q  &  N   &3.168 &1.7$\times10^{-4}$& 1.4000& 25.38    & 3.34 & 73.6     & 1	       & 211 &205  &    AL634$\ast$   \\
0838$+$133    &   0.681  &  18.15 &  Q  & ...  &0.287 &8.0$\times10^{-5}$& 1.4250& 12.93    & 0.35 & 2216.6   & 2hs	       & 76  &94   &    AH480         \\
0851$+$202    &   0.306  &  15.43 &  B  &  Y   &1.563 &1.4$\times10^{-4}$& 1.5315& 15.17    & 1.57 & 10.7     & 1hs, 1          & 260 &251  &    AS764         \\
0906$+$015    &   1.024  &  17.31 &  Q  &  Y   &1.00  &0.4$\times10^{-3}$& 1.4650& 20.66    & 1.00 & 38.0     & 1hs	       & 46  &54   &    Murphy        \\
0917$+$624    &   1.446  &  19.50 &  Q  & ...  &1.11  &0.4$\times10^{-3}$& 1.4650& 15.57    & 1.11 & 6.4      & 1hs+2           & 341 &234  &    Murphy        \\
0923$+$392    &   0.695  &  17.03 &  Q  &  N   &2.399 &3.1$\times10^{-4}$& 1.5524& 4.29     & 2.83 & 361.8    & 1hs+2           & 98  &83   &    AB310         \\
0945$+$408    &   1.249  &  18.05 &  Q  &  N   &1.23  &0.9$\times10^{-3}$& 1.4650& 18.60    & 1.23 & 95.0     & 1hs	       & 114 &32   &    Murphy        \\
0955$+$476$^c$&   1.882  &  18.65 &  Q  & ...  &0.606 &1.0$\times10^{-4}$& 1.4899& 2.48	    & 0.62 & 1.1      & c              & 124 &270  &    AP188         \\
1036$+$054    &   0.473  &   .... &  Q  & ...  &0.916 &4.9$\times10^{-5}$& 1.4250& 6.14     & 0.94 & 57.0     & 1hs+2           & 350 &226  &    AC874$\dagger$\\
1038$+$064    &   1.265  &  16.70 &  Q  & ...  &1.465 &1.5$\times10^{-4}$& 1.6649& 11.86    & 1.49 & 11.0     & 1 	       & 146 &166  &    GX007A        \\
1045$-$188    &   0.595  &  18.20 &  Q  & ...  &0.726 &6.6$\times10^{-5}$& 1.4250& 8.57     & 0.76 & 509.4    & 1hs+2           & 146 &125  &    AC874$\dagger$\\
1055$+$018    &    0.89  &  18.28 &  Q  &  Y   &2.675 &1.5$\times10^{-4}$& 1.4250& 10.99    & 2.70 & 230.8    & 2hs	       & 300 &186  &    AB631         \\
1124$-$186    &   1.048  &  18.65 &  Q  & ...  &0.652 &1.9$\times10^{-4}$& 1.4250& ....	    & 0.66 & 12.3     & 1hs	       & 170 &134  &    AD337         \\
1127$-$145    &   1.184  &  16.90 &  Q  & ...  &4.460 &2.3$\times10^{-4}$& 1.4649& 14.17    & 4.58 & 59.3     & 1hs             & 66  &44   &    AB379         \\
1150$+$812    &    1.25  &  19.40 &  Q  & ...  &1.874 &6.5$\times10^{-4}$& 1.4000& 7.08     & 1.89 & 89.2     & 1hs+2, 2        & 163 &257  &    AL634$\ast$   \\
1156$+$295    &   0.729  &  14.41 &  Q  &  Y   &1.433 &2.4$\times10^{-4}$& 1.4899& 24.85    & 1.55 & 196.1    & 2, 1hs+2        & 14  &344  &    AA025         \\
1213$-$172    &    ....  &  21.40 &  U  & ...  &1.676 &1.0$\times10^{-4}$& 1.4250& ....     & 1.85 & 119.3    & 1	       & 117 &266  &    AC874$\dagger$\\
1219$+$044    &   0.965  &  17.98 &  Q  & ...  &0.548 &5.9$\times10^{-5}$& 1.4250& 2.34	    & 0.60 & 155.5    & 1hs+2           & 172 &0    &    AC874$\dagger$\\
1222$+$216    &   0.432  &  17.50 &  Q  & ...  &0.930 &8.0$\times10^{-5}$& 1.4000& 21.02    & 1.10 & 956.4    & 2hs	       & 351 &76   &    AL634$\ast$   \\
1226$+$023    &   0.158  &  12.85 &  Q  &  N   &34.26 &1.5$\times10^{-3}$& 1.3659& 13.44    & 34.89& 17671    & 1hs	       & 238 &222  &    AE104         \\
1228$+$126    &  0.0044  &  12.86 &  G  & ...  &3.558 &2.3$\times10^{-4}$& 1.6649& 0.032    & 3.88 & 115012   & 2	       & 291 &67   &    BC079         \\
1253$-$055    &   0.536  &  17.75 &  Q  &  Y   &10.17 &6.4$\times10^{-4}$& 1.6649& 20.58    & 10.56& 2095.0   & 2hs	       & 245 &201  &    W088D5        \\
1308$+$326    &   0.996  &  15.24 &  Q  &  Y   &1.321 &1.5$\times10^{-4}$& 1.4000& 27.14    & 1.33 & 69.1     & 2hs, 1hs+2       & 284 &359  &    AL634$\ast$   \\
1324$+$224    &     1.4  &  18.90 &  Q  & ...  &1.128 &9.6$\times10^{-5}$& 1.4000& ....     & 1.14 & 20.4     & 1hs             & 343 &237  &    AL634$\ast$   \\
1334$-$127    &   0.539  &  19.00 &  Q  &  Y   &1.992 &9.9$\times10^{-5}$& 1.4899& 10.26    & 2.07 & 151.0    & 1hs, 1hs+2       & 147 &106  &    AD176         \\
1413$+$135    &   0.247  &  20.50 &  B  & ...  &1.056 &1.1$\times10^{-4}$& 1.4899& 1.79     & 1.08 & 5.8      & 2	       & 55  &256  &    AC301         \\
1417$+$385    &   1.831  &  19.69 &  Q  & ...  &0.506 &6.4$\times10^{-5}$& 1.4000& 15.44    & 0.52 & 2.5      & 1              & 164 &123  &    AL634$\ast$   \\
1458$+$718    &   0.905  &  16.78 &  Q  &  N   &5.760 &8.7$\times10^{-4}$& 1.4000& 7.05	    & 7.57 & 68.9     & c, 1           & 164 &11   &    AL634$\ast$   \\
1502$+$106    &   1.839  &  18.56 &  Q  &  Y   &1.795 &7.9$\times10^{-5}$& 1.4000& 14.77    & 1.82 & 38.3     & 1hs	       & 116 &159  &    AL634$\ast$   \\
1504$-$166    &   0.876  &  18.50 &  Q  &  Y   &2.354 &4.0$\times10^{-4}$& 1.4899& 4.31     & 2.39 & 11.4     & 1hs, 2hs         & 165 &164  &    AY012         \\
1510$-$089    &    0.36  &  16.54 &  Q  &  Y   &1.380 &1.0$\times10^{-4}$& 1.4899& 20.15    & 1.45 & 180.2    & 1hs, 2hs         & 328 &163  &    AY012         \\
1538$+$149    &   0.605  &  17.30 &  B  &  Y   &1.479 &6.0$\times10^{-5}$& 1.4000& 8.73     & 1.67 & 71.4     & ch, 1          & 323 &320  &    AL634$\ast$   \\
1546$+$027    &   0.414  &  17.45 &  Q  &  Y   &1.15  &0.4$\times10^{-3}$& 1.4650& 12.07    & 1.15 & 18.8     & 1hs+2           & 170 &162  &    Murphy        \\
1548$+$056    &   1.422  &  19.50 &  Q  &  Y   &2.106 &1.6$\times10^{-4}$& 1.5524& 11.56    & 2.21 & 42.9     & ch	       & 8   &.... &    AB310         \\
1606$+$106    &   1.226  &  18.70 &  Q  & ...  &1.35  &1.3$\times10^{-3}$& 1.4650& 18.90    & 1.35 & 26.5     & 1hs	       & 332 &271  &    Murphy        \\
1611$+$343    &   1.397  &  18.11 &  Q  &  N   &2.83  &0.5$\times10^{-3}$& 1.4650& 14.09    & 2.83 & 20.6     & 2hs	       & 166 &194  &    Murphy        \\
1633$+$382    &   1.814  &  18.00 &  Q  &  Y   &2.17  &0.4$\times10^{-3}$& 1.4650& 29.46    & 2.17 & 32.0     & 1hs+2           & 284 &176  &    Murphy        \\
1637$+$574    &   0.751  &  16.90 &  Q  &  N   &0.996 &1.1$\times10^{-4}$& 1.5524& 10.61    & 1.01 & 71.2     & 1hs	       & 198 &282  &    AB310         \\
1638$+$398    &   1.666  &  19.37 &  Q  & ...  &1.147 &2.0$\times10^{-4}$& 1.4899& 12.27    & 1.17 & 27.6     & 2hs	       & 300 &150  &    AR250         \\
1641$+$399    &   0.593  &  16.62 &  Q  &  Y   &7.153 &1.3$\times10^{-3}$& 1.5149& 19.27    & 7.95 & 1476.9   & 1hs+2, 2hs       & 288 &327  &    AS396         \\
1655$+$077    &   0.621  &  20.00 &  Q  &  Y   &1.199 &1.4$\times10^{-4}$& 1.5524& 14.44    & 1.27 & 199.1    & 1hs+2           & 314 &269  &    AB310         \\
1726$+$455    &   0.717  &  18.10 &  Q  & ...  &0.993 &1.3$\times10^{-4}$& 1.4899& 1.81     & 1.00 & 55.3     & 1hs+2           & 271 &328  &    AK139         \\
1730$-$130    &   0.902  &  19.50 &  Q  & ...  &6.101 &3.4$\times10^{-4}$& 1.4250& 35.69    & 6.13 & 517.8    & 2hs, 1hs+2       & 2   &273  &    AL269         \\
1739$+$522    &   1.375  &  18.70 &  Q  &  Y   &1.571 &1.5$\times10^{-4}$& 1.4899& ....     & 1.61 & 27.6     & 1hs+2           & 15  &262  &    AC301         \\
1741$-$038$^c$&   1.054  &  20.40 &  Q  &  Y   &1.677 &1.5$\times10^{-4}$& 1.4250& ....     & 1.70 & 3.5      & c	       & 201 &134  &    AL269         \\
1749$+$096$^c$&   0.322  &  16.78 &  B  &  Y   &1.039 &1.6$\times10^{-4}$& 1.4899& 6.84     & 1.05 & 4.9      & c              & 38  &.... &    AC301         \\
1751$+$288    &   1.118  &  19.60 &  Q  & ...  &0.265 &1.7$\times10^{-4}$& 1.4350& 3.07     & 0.27 & 8.1      & 2, 1           & 349 &40   &    AS659         \\
1758$+$388    &   2.092  &  17.98 &  Q  & ...  &0.327 &1.7$\times10^{-4}$& 1.5149& 2.38	    & 0.33 & 3.2      & 1hs+2           & 266 &270  &    AS396         \\
1800$+$440    &   0.663  &  17.90 &  Q  & ...  &0.465 &2.1$\times10^{-4}$& 1.5149& 15.41    & 0.50 & 246.6    & 1hs+2           & 203 &239  &    AS396         \\
1803$+$784    &    0.68  &  15.90 &  B  &  Y   &1.969 &1.0$\times10^{-4}$& 1.4000& 8.97     & 1.98 & 20.8     & 2hs             & 266 &192  &    AL634$\ast$   \\
1807$+$698    &   0.051  &  14.22 &  B  &  Y   &1.133 &1.1$\times10^{-4}$& 1.5149& 0.10     & 1.20 & 368.7    & 1hs	       & 261 &242  &    AB700         \\
1823$+$568    &   0.664  &  19.30 &  B  &  Y   &0.873 &6.1$\times10^{-4}$& 1.4250& 20.85    & 0.95 & 137.4    & 2 	       & 201 &94   &    AM672         \\
1828$+$487    &   0.692  &  16.81 &  Q  &  N   &5.140 &6.3$\times10^{-4}$& 1.5524& 13.65    & 8.21 & 5431.2   & 2hs	       & 310 &119  &    AB310         \\
1849$+$670    &   0.657  &  16.90 &  Q  & ...  &0.468 &1.1$\times10^{-4}$& 1.4000& 30.63    & 0.47 & 101.0    & 1hs+2           & 308 &209  &    AL634$\ast$   \\
1928$+$738    &   0.302  &  16.06 &  Q  &  N   &3.186 &1.6$\times10^{-4}$& 1.4000& 8.43     & 3.22 & 356.5    & 1hs+2, 2hs       & 164 &165  &    AL634$\ast$   \\
1936$-$155    &   1.657  &  20.30 &  Q  &  Y   &1.064 &1.8$\times10^{-4}$& 1.4899& 2.59     & 1.08 & 10.5     & 1hs+2, 2        & 180 &214  &    AD167         \\
1957$+$405    &  0.0561  &  15.10 &  G  & ...  &33.56$^\ddagger$&6.5$\times10^{-3}$& 1.5245& 0.21& 0.65 & 1.4$\times10^6$ & 2hs & 282 &291  &    AC166         \\
1958$-$179$^c$&    0.65  &  18.60 &  Q  &  Y   &1.789 &1.5$\times10^{-4}$& 1.4899& 1.89	    & 1.82 & 9.4      & c	       & 207 &.... &    AA072         \\
2005$+$403    &   1.736  &  19.00 &  Q  & ...  &2.420 &5.1$\times10^{-4}$& 1.6657& 12.21    & 2.47 & 10.6     & 1hs+2           & 127 &137  &    AK4010        \\
2008$-$159    &    1.18  &  18.30 &  Q  & ...  &0.546 &1.0$\times10^{-4}$& 1.4899& 7.98     & 0.55 & 7.5      & 1hs	       & 7   &5    &    AK240         \\
2021$+$317    &    ....  &   .... &  U  & ...  &2.968 &6.8$\times10^{-4}$& 1.6657& ....     & 3.06 & 132.9    & ch, 1          & 201 &.... &    AK4010        \\
2021$+$614    &   0.227  &  19.00 &  G  &  N   &2.647 &2.9$\times10^{-4}$& 1.4899& 0.41     & 2.67 & 1.3      & 2	       & 33  &.... &    AM178         \\
2037$+$511    &   1.686  &  21.00 &  Q  & ...  &4.604 &6.4$\times10^{-4}$& 1.6657& 3.29	    & 4.97 & 657.6    & 1hs	       & 214 &327  &    AK4010        \\
2121$+$053    &   1.941  &  20.40 &  Q  &  Y   &1.08  &0.6$\times10^{-3}$& 1.4650& 13.28    & 1.08 & 4.8      & 1, 1hs          & 271 &.... &    Murphy        \\
2128$-$123    &   0.501  &  16.11 &  Q  &  N   &1.393 &1.3$\times10^{-4}$& 1.4899& 6.94     & 1.42 & 39.8     & 1hs+2           & 209 &190  &    AY012         \\
2131$-$021    &   1.285  &  19.00 &  B  &  Y   &1.320 &1.3$\times10^{-4}$& 1.5149& 20.03    & 1.37 & 151.9    & 2hs	       & 108 &137  &    AB700         \\
2134$+$004    &   1.932  &  17.11 &  Q  &  N   &4.82  &0.4$\times10^{-3}$& 1.4650& 5.93     & 4.82 & 6.6      & 1hs	       & 266 &298  &    Murphy        \\
2136$+$141$^c$&   2.427  &  18.90 &  Q  & ...  &1.131 &1.4$\times10^{-4}$& 1.5524& 5.43	    & 1.14 & 0.8      & c	       & 205 &.... &    AB310         \\
2145$+$067    &    0.99  &  16.47 &  Q  &  N   &2.840 &1.7$\times10^{-4}$& 1.4000& 2.49     & 2.87 & 27.7     & 1hs, 1          & 119 &313  &    AL634$\ast$   \\
2155$-$152    &   0.672  &  18.30 &  Q  &  Y   &2.627 &1.5$\times10^{-4}$& 1.4000& 18.11    & 2.70 & 304.7    & 1hs+2           & 214 &195  &    AL634$\ast$   \\
2200$+$420    &  0.0686  &  14.72 &  B  &  Y   &1.970 &7.0$\times10^{-5}$& 1.4000& 10.57    & 1.99 & 14.2     & 2, 1hs+2        & 182 &143  &    AL634$\ast$   \\
2201$+$171    &   1.075  &  19.50 &  Q  & ...  &0.820 &5.8$\times10^{-5}$& 1.4000& 2.54     & 0.87 & 74.6     & 1hs, 2hs         & 45  &267  &    AL634$\ast$   \\
2201$+$315    &   0.295  &  15.58 &  Q  &  N   &1.517 &1.3$\times10^{-4}$& 1.4000& 7.87	    & 1.54 & 378.4    & 2hs	       & 219 &251  &    AL634$\ast$   \\
2209$+$236$^c$&   1.125  &  20.66 &  Q  & ...  &0.425 &4.9$\times10^{-5}$& 1.4000& 3.43     & 0.43 & 0.9      & c              & 54  &.... &    AL634$\ast$   \\
2216$-$038    &   0.901  &  16.38 &  Q  & ...  &1.722 &1.1$\times10^{-4}$& 1.4000& 5.62     & 1.76 & 312.7    & 2hs	       & 190 &143  &    AL634$\ast$   \\
2223$-$052    &   1.404  &  18.39 &  Q  &  Y   &6.769 &3.4$\times10^{-4}$& 1.4000& 17.33    & 7.13 & 91.6     & 1	       & 99  &335  &    AL634$\ast$   \\
2227$-$088    &    1.56  &  17.43 &  Q  &  Y   &0.920 &4.2$\times10^{-5}$& 1.4000& 8.14     & 0.93 & 8.4      & 1hs	       & 348 &314  &    AL634$\ast$   \\
2230$+$114    &   1.037  &  17.33 &  Q  &  Y   &6.633 &3.7$\times10^{-4}$& 1.4000& 15.41    & 6.99 & 148.0    & 1	       & 147 &144  &    AL634$\ast$   \\
2243$-$123    &   0.632  &  16.45 &  Q  &  Y   &2.247 &8.4$\times10^{-5}$& 1.4000& 5.49     & 2.27 & 27.7     & 1hs	       & 29  &44   &    AL634$\ast$   \\
2251$+$158    &   0.859  &  16.10 &  Q  &  Y   &13.88 &4.9$\times10^{-4}$& 1.4000& 14.19    & 14.09& 822.0    & 1hs	       & 297 &311  &    AL634$\ast$   \\
2331$+$073    &   0.401  &  16.04 &  Q  & ...  &0.602 &4.3$\times10^{-5}$& 1.4000& 4.47     & 0.61 & 38.4     & 1hs+2           & 225 &251  &    AL634$\ast$   \\
2345$-$167    &   0.576  &  18.41 &  Q  &  Y   &1.921 &8.1$\times10^{-5}$& 1.4000& 13.45    & 1.99 & 142.7    & 1hs	       & 141 &227  &    AL634$\ast$   \\
2351$+$456    &   1.986  &  20.60 &  Q  & ...  &2.260 &6.3$\times10^{-5}$& 1.4000& 27.09    & 2.35 & 6.9      & c, 1hs          & 278 &260  &    AL634$\ast$   \\
\enddata
\tablecomments
{{\scriptsize Col.~1 \& 2: IAU B1950 names of the MOJAVE sources and their redshifts.
$c$ = sources do not show any discernable extended emission.
Col.~3: Apparent V-band magnitude obtained from the AGN catalog of 
\citet{Veron06} or NED. 
Col.~4: Blazar classification $-$ Q = quasar, B = BL~Lac object, G = radio
galaxy, U = Unidentified.
Col.~5: Highly optically polarized $-$ Y = Yes, N = No, .... = no data
available.
Col.~6 \& 7: Peak and $rms$ intensity in radio map, respectively.
$I_{rms}$ for the sources in \citet{Murphy93} are the CLEV values in their Table 2.
$\ddagger$ for Cygnus A, the peak intensity corresponds to the south-eastern
hot spot.
Col.~8: Central observing frequency in GHz.
Col.~9: Apparent parsec-scale speed of the fastest jet feature. 
Col.~10: Integrated core flux density at $\sim$1.4 GHz obtained with JMFIT.
Col.~11: Extended flux density obtained by subtracting the core
from the total radio flux density.
Col.~12: Radio morphology $-$ c = core only, ch = core-halo, 1hs = 1 hot spot, 
2hs = 2 hot spots, 1 = 1 sided, 2 = 2 sided, 1hs+2 = 2 sided structure with 
hot spot on one side. Alternate morphologies are listed
separated by commas.
Col.~13: Parsec-scale jet position angle.
Col.~14: Kiloparsec-scale jet position angle.
Col.~15: References and VLA archive Project IDs for the radio data $-$
$\dagger$=New data presented in this paper.
$\ast$=Data presented in \citet{Cooper07}. Murphy = \citet{Murphy93}.}}
\label{tabsample}
\end{deluxetable}

\begin{deluxetable}{llllllll}
\tabletypesize{\scriptsize}
\tablecaption{Correlation Results}
\tablewidth{0pt}
\tablehead{
\colhead{Param 1} & \colhead{Param 2} & \colhead{Class}& \colhead{Spearman} &
\colhead{Spearman}& \colhead{Kendall $\tau$} & \colhead{Kendall $\tau$} &
\colhead{Correl ?}\\
\colhead{} & \colhead{} &\colhead{} & \colhead{Coeff} & \colhead{Prob} & 
\colhead{Coeff}& \colhead{Prob}& \colhead{} \\
\colhead{(1)}  & \colhead{(2)} & \colhead{(3)} & \colhead{(4)} & 
\colhead{(5)}  & \colhead{(6)} & \colhead{(7)} & \colhead{(8)}}
\startdata
$L_{core}$  & $L_{ext}$        &ALL& 0.61   & 1.6$\times10^{-13}$ &0.45    &$<1\times10^{-5}$ & YES  \\
$M_{abs}$   & $L_{ext}$        &ALL&$-$0.39 & 9.8$\times10^{-6}$ &$-$0.27 &1.1$\times10^{-6}$ & YES  \\
$L_{ext}$   & $\beta_{app}$    &ALL& 0.44   & 5.2$\times10^{-7}$ & 0.31   &5.3$\times10^{-7}$ & YES  \\
            &                  &Q&0.36&0.0004&0.24&0.0005& YES \\
            &                  &B&0.32&0.193&0.24&0.148&NO \\
$M_{abs}$   & $\beta_{app}$    &ALL&$-$0.27 & 0.0025&$-$0.18&0.0031& YES\\
$R_c$       & $L_{core}$       &ALL&0.26 & 0.003   &0.19    & 0.002  & YES  \\
            &                  &Q& 0.39  & 8.5$\times10^{-5}$ & 0.27   &8.8$\times10^{-5}$ & YES  \\
            &                  &B&$-$0.02& 0.912   &$-$0.01 &0.909   & NO   \\
$R_c$       & $L_{ext}$        &ALL&$-$0.49 & 1.5$\times10^{-8}$ &$-$0.35 &$<1\times10^{-5}$ & YES  \\
$R_c$       & $\beta_{app}$    &ALL&$-$0.18&0.046&$-$0.12&0.044&YES?\\
            &                  &Q&$-$0.28& 0.006   &$-$0.19 & 0.006  & YES  \\
            &                  &B&0.02   & 0.922   &0.01    & 0.939  & NO   \\
$R_c$       & $\Delta$PA       &ALL&$-$0.13& 0.146   &$-$0.09 & 0.142  & NO  \\
            &                  &Q&$-$0.18& 0.087   &$-$0.12 & 0.082  & YES? \\
            &                  &B&$-$0.15& 0.579   &$-$0.10 & 0.589  & NO   \\
$R_v$       & $L_{core}$       &ALL& 0.62   & 6.5$\times10^{-15}$ & 0.45   &$<1\times10^{-5}$ & YES  \\
$R_v$       & $\beta_{app}$    &ALL& 0.11   & 0.200   & 0.07   &0.226   & NO   \\
$R_v$       & $\Delta$PA       &ALL&0.23&0.011&0.15&0.017&YES?\\
            &                  &Q& 0.26  & 0.011   & 0.17   &0.015   & YES? \\
            &                  &B& 0.38  & 0.143   & 0.23   &0.207   & NO   \\ 
            &                  &FRII&0.24&0.029&0.16&0.032&YES?\\
            &                  &FRI/II&0.18&0.277&0.11&0.313&NO\\
            &                  &FRI&0.12&0.658&0.07&0.701&NO\\
$\Delta$PA  & $z$              &ALL& 0.06   & 0.514    & 0.04  & 0.498  & NO \\
$L_{ext}/M_{abs}$&$z$          &Q+B&0.21& 0.023&0.14&0.021&YES?\\
                 &             &Q& 0.27  & 0.007   & 0.19   &0.007   & YES  \\
            &                  &B& 0.31  & 0.209   & 0.16   &0.343   & NO   \\ 
$L_{ext}/M_{abs}$&$\beta_{app}$&Q+B&$-$0.02 & 0.810   &$-$0.01 &0.768   & NO \\
$L_{ext}/M_{abs}$&$\Delta$PA   &Q+B&$-$0.34 & 0.0003  &$-$0.23 & 0.0005 & YES\\
\enddata
\tablecomments{
Cols.~1 \& 2: Parameters being examined for correlations. Results from a 
partial regression analysis are cited in the main text.
The ten core-only sources have been excluded from correlations with
extended luminosity. Including the upper limits in extended luminosity as
detections, does not alter any of the observed trends.
Col.~3: The correlation test results have been presented separately
if they were different for quasars and BL Lacs considered alone.
ALL = quasars + BL Lacs + radio galaxies, Q = only quasars,
B = only BL Lacs, Q+B = quasars + BL Lacs. FRII, FRI/II, and FRI are as defined
in \S4.5.
Cols.~4 \& 5: Spearman rank correlation coefficient and 
chance probability. Cols.~6 \& 7: Kendall Tau correlation coefficient and
chance probability. Col.~8: Indicates if the parameters are significantly 
correlated. `YES?' indicates a marginal correlation.
We note that at the 95\% confidence level, 1.7 spurious correlations could
arise from the $\approx$35 correlations that we tested.} 
\label{tabcorrel}
\end{deluxetable}

\acknowledgments
We would like to thank the anonymous referee for a careful assessment
of the manuscript, which has led to significant improvement.
PK would like to thank John Peterson for useful discussions on galaxy clusters,
and Chris O'Dea for insightful suggestions. 
The MOJAVE project is supported under National Science Foundation grant 
0807860-AST and NASA-Fermi grant NNX08AV67G. 
This research has made use of the NASA/IPAC Extragalactic Database (NED) which 
is operated by the Jet Propulsion Laboratory, California Institute of 
Technology, under contract with the National Aeronautics and Space Administration.
The National Radio Astronomy Observatory is a facility of the National 
Science Foundation operated under cooperative agreement by Associated 
Universities, Inc.


\appendix
\section{PREVIOUSLY UNPUBLISHED RADIO IMAGES FROM THE ARCHIVE}

\begin{figure}[ht]
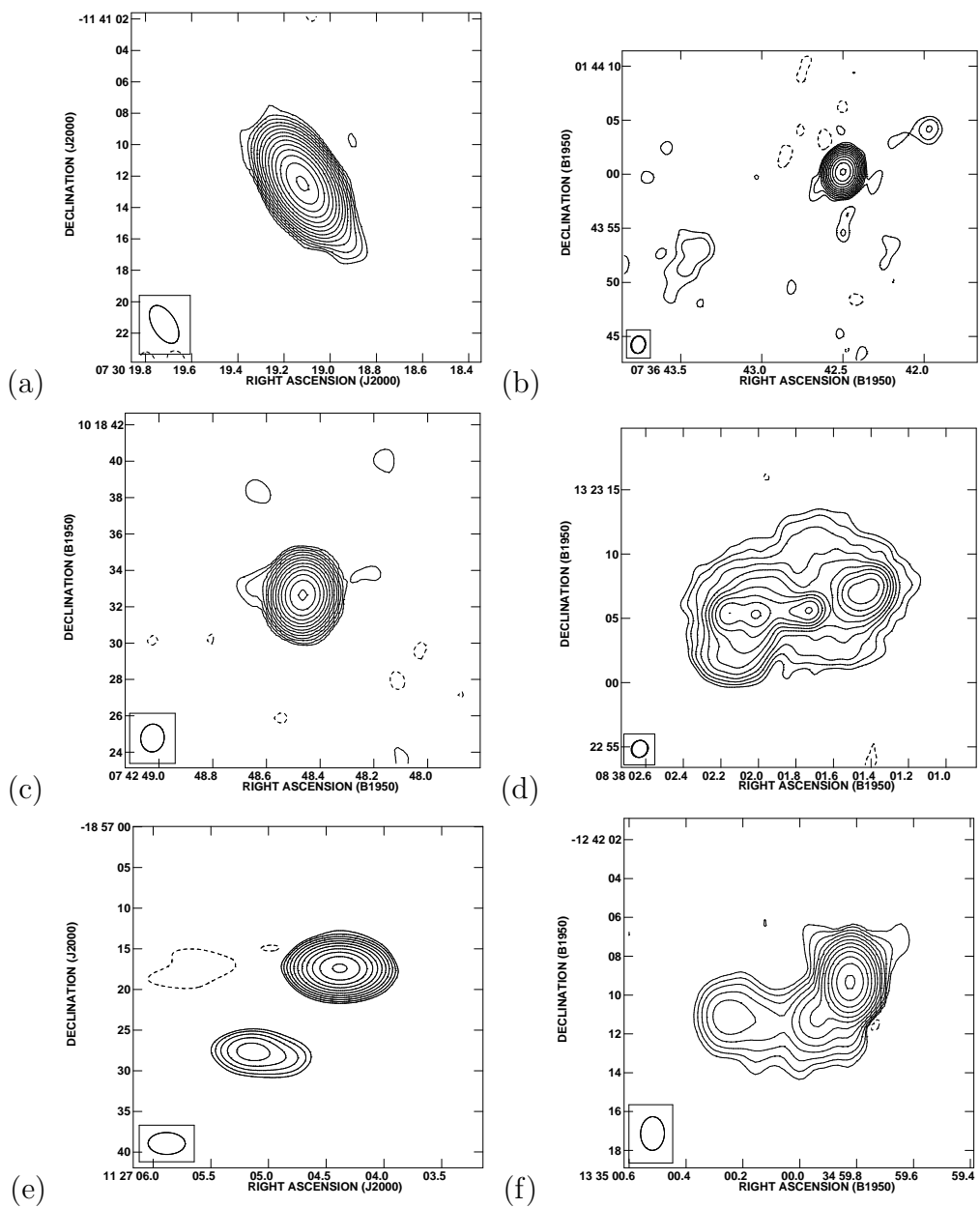

\centering{
(a)
\includegraphics[width=6cm]{figure13a.ps}
(b)
\includegraphics[width=6cm]{figure13b.ps}

(c)
\includegraphics[width=6cm]{figure13c.ps}
(d)
\includegraphics[width=6cm]{figure13d.ps}

(e)
\includegraphics[width=6cm]{figure13e.ps}
(f)
\includegraphics[width=6cm]{figure13f.ps}}
\caption{\small New radio images from the VLA archive of 
(a) 0727$-$115, (b) 0736+017, (c) 0742+103, (d) 0838+133, (e) 1124$-$186, and
(f) 1334$-$127. The contours are in percentage of the peak surface brightness and 
increase in steps of 2. The lowest contour levels and peak surface brightness are, 
(a) $\pm$0.01, 3.15 Jy~beam$^{-1}$;
(b) $\pm$0.042, 2.33 Jy~beam$^{-1}$;
(c) $\pm$0.021, 3.61 Jy~beam$^{-1}$;
(d) $\pm$0.17, 289.3 mJy~beam$^{-1}$;
(e) $\pm$0.085, 651.2 mJy~beam$^{-1}$;
(f) $\pm$0.042, 1.94 Jy~beam$^{-1}$.}
\end{figure}

\begin{figure}[ht]
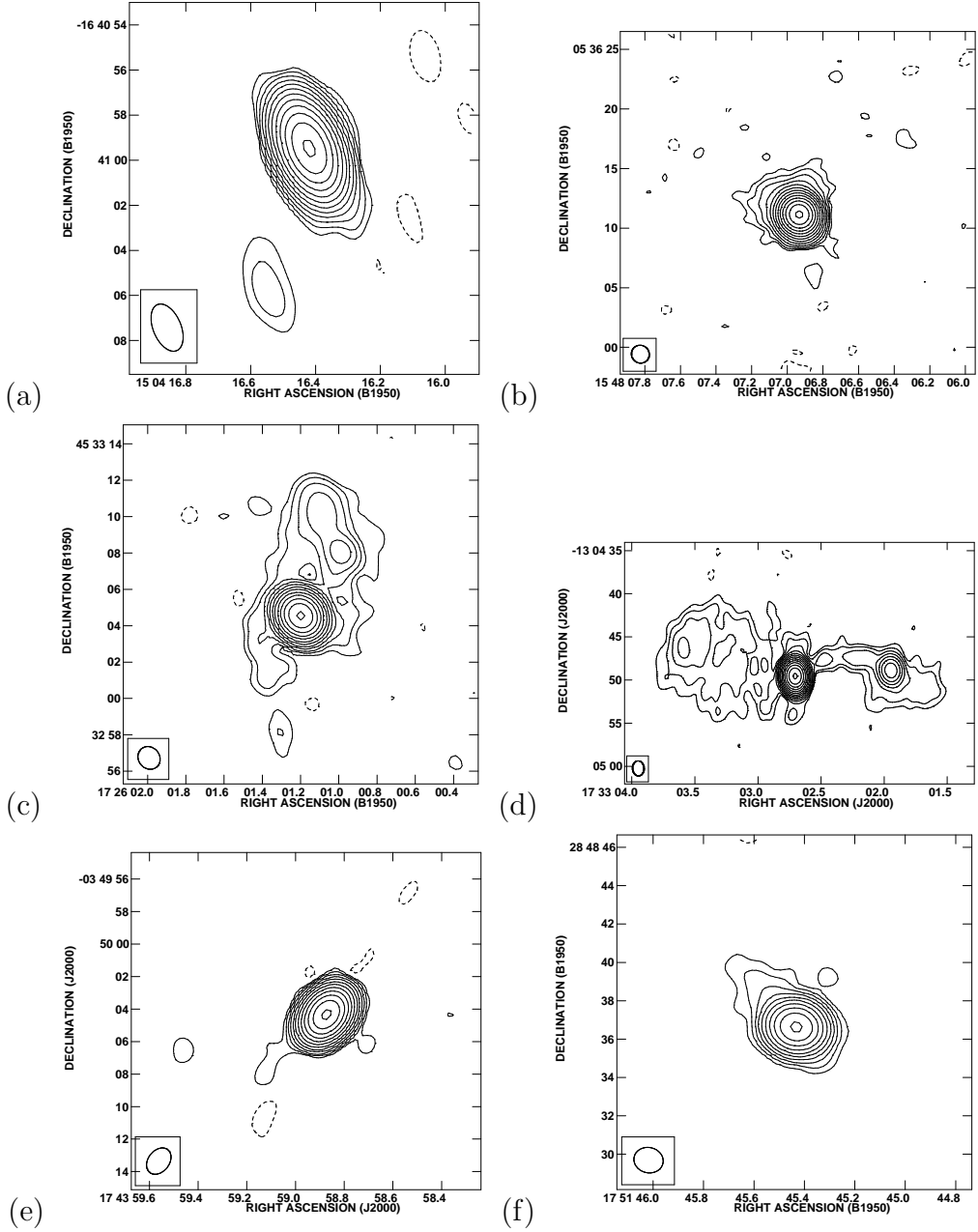

\centering{
(a)
\includegraphics[width=6cm]{figure14a.ps}
(b)
\includegraphics[width=6cm]{figure14b.ps}

(c)
\includegraphics[width=6cm]{figure14c.ps}
(d)
\includegraphics[width=6cm]{figure14d.ps}

(e)
\includegraphics[width=6cm]{figure14e.ps}
(f)
\includegraphics[width=6cm]{figure14f.ps}}
\caption{\small New radio images from the VLA archive of 
(a) 1504$-$166, (b) 1548+056, (c) 1726+455, (d) 1733$-$130, (e) 1741$-$038, and
(f) 1751+288. The contours are in percentage of the peak surface brightness and 
increase in steps of 2. The lowest contour levels and peak surface brightness are, 
(a) $\pm$0.042, 2.36 Jy~beam$^{-1}$;
(b) $\pm$0.021, 2.09 Jy~beam$^{-1}$;
(c) $\pm$0.042, 998.9 mJy~beam$^{-1}$;
(d) $\pm$0.021, 6.11 Jy~beam$^{-1}$;
(e) $\pm$0.042, 1.68 Jy~beam$^{-1}$;
(f) $\pm$0.17, 263.5 mJy~beam$^{-1}$.}
\end{figure}

\begin{figure}[ht]
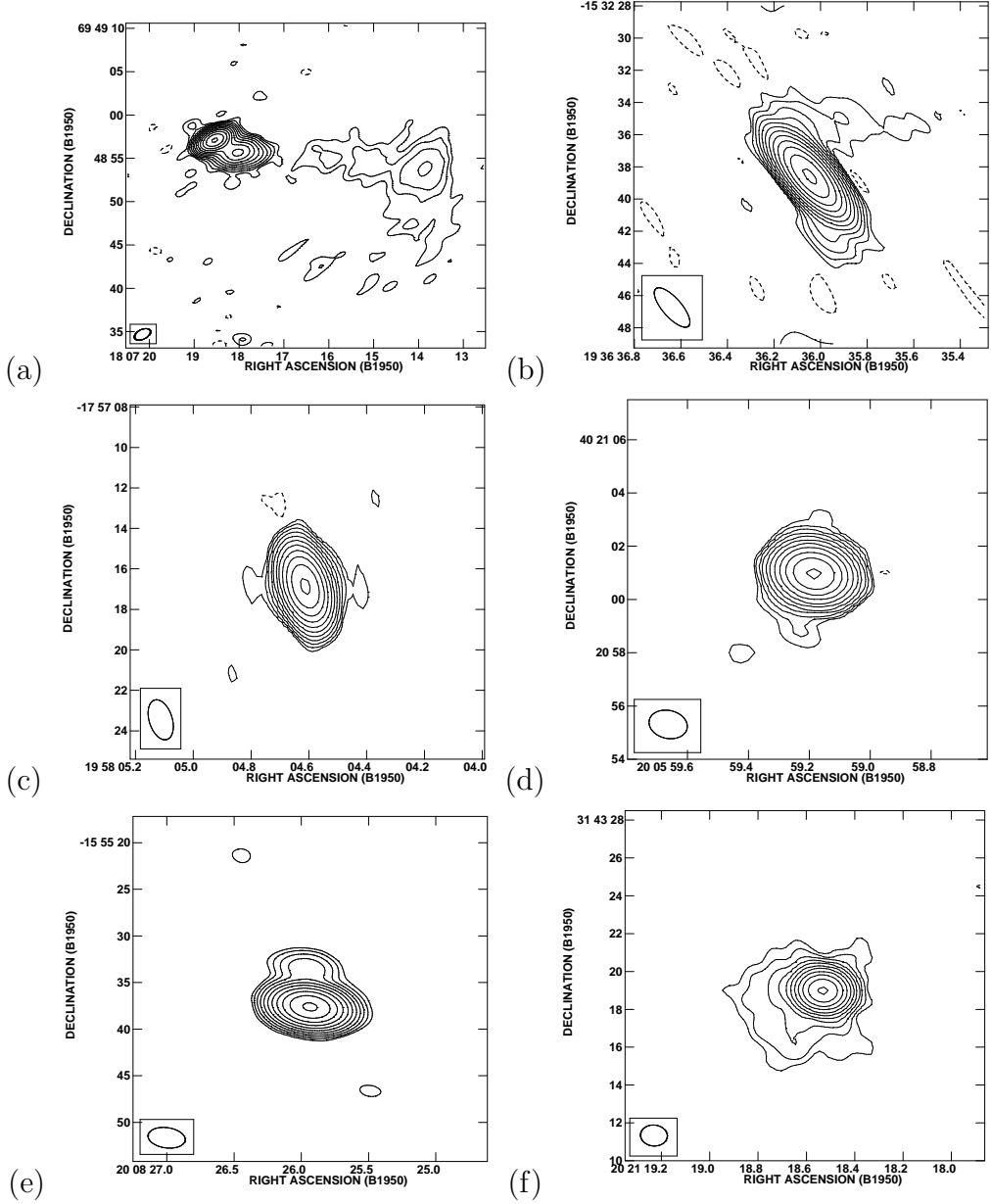

\centering{
(a)
\includegraphics[width=6cm]{figure15a.ps}
(b)
\includegraphics[width=6cm]{figure15b.ps}

(c)
\includegraphics[width=6cm]{figure15c.ps}
(d)
\includegraphics[width=6cm]{figure15d.ps}

(e)
\includegraphics[width=6cm]{figure15e.ps}
(f)
\includegraphics[width=6cm]{figure15f.ps}}
\caption{\small New radio images from the VLA archive of 
(a) 1807+698, (b) 1936$-$155, (c) 1958$-$179, (d) 2005+403, (e) 2008$-$159, and 
(f) 2021+317. The contours are in percentage of the peak surface brightness and 
increase in steps of 2. The lowest contour levels and peak surface brightness are, 
(a) $\pm$0.042, 1.14 Jy~beam$^{-1}$;
(b) $\pm$0.085, 1.05 Jy~beam$^{-1}$;
(c) $\pm$0.085, 1.77 Jy~beam$^{-1}$;
(d) $\pm$0.085, 2.43 Jy~beam$^{-1}$;
(e) $\pm$0.085, 544.5 mJy~beam$^{-1}$;
(f) $\pm$0.085, 2.97 Jy~beam$^{-1}$.}
\end{figure}

\begin{figure}[ht]
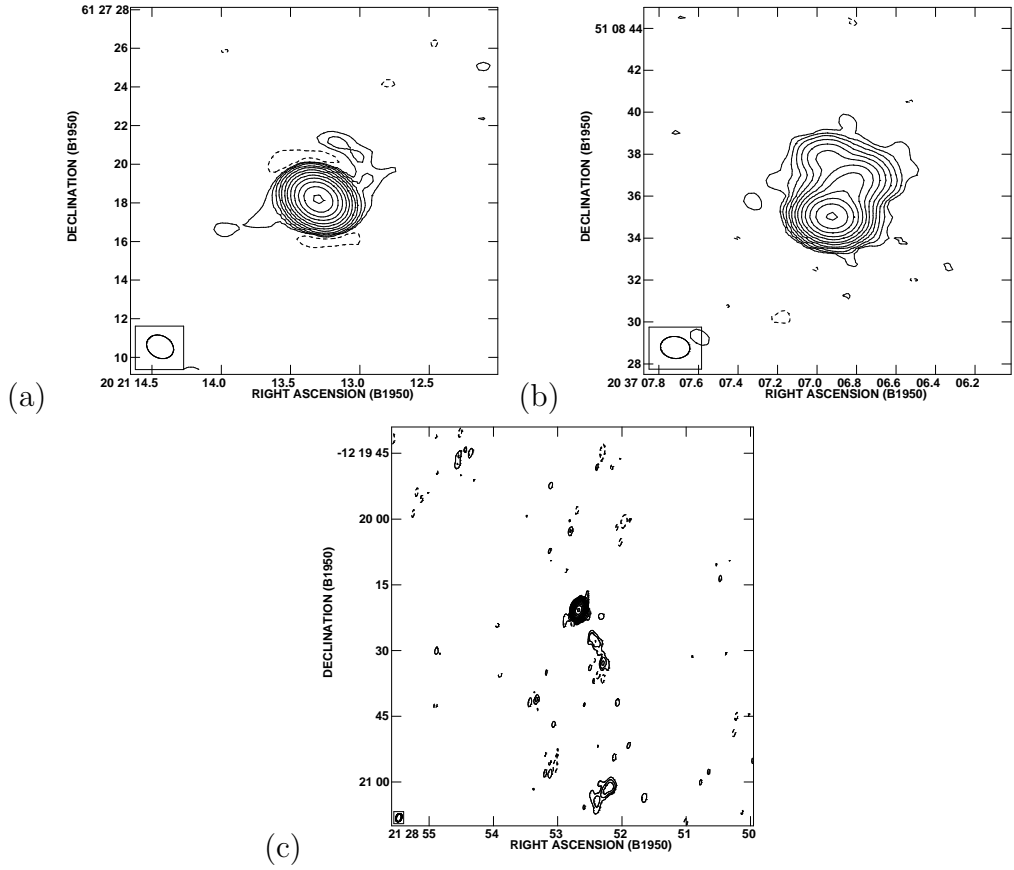

\centering{
(a)
\includegraphics[width=6cm]{figure16a.ps}
(b)
\includegraphics[width=6cm]{figure16b.ps}

(c)
\includegraphics[width=6cm]{figure16c.ps}}
\caption{\small New radio images from the VLA archive of 
(a) 2021+614, (b) 2037+511, and (c) 2128$-$123. The contours are in percentage of 
the peak surface brightness and increase in steps of 2. The lowest contour levels 
and peak surface brightness are,
(a) $\pm$0.042, 2.64 Jy~beam$^{-1}$;
(b) $\pm$0.042, 4.61 Jy~beam$^{-1}$;
(c) $\pm$0.042, 1.37 Jy~beam$^{-1}$.}
\end{figure}

\end{document}